\documentclass[superscriptaddress,groupedaddress,nofootnoteinbib,11pt]{article}
\pdfoutput=1 
\usepackage{jheppub}

\usepackage{mathtools,slashed,mathrsfs}
\usepackage[caption=false]{subfig}
\usepackage{dcolumn}
\usepackage{multirow}
\usepackage{tabularx}
\usepackage{booktabs}
\usepackage{bm}
\usepackage{amsmath}
\usepackage{comment}
\usepackage{setspace}
\usepackage[dvipsnames]{xcolor}
\usepackage[normalem]{ulem} 
\usepackage{enumerate}
\usepackage{siunitx}
\usepackage{float}
\usepackage{hyperref}

\usepackage{physics}

\newcommand{\NDW}{N_{\text{DW}}}

\newcommand{\tann}{t_{\text{ann}}}

\newcommand{\tEMD}{t_{\text{EMD}}}
\newcommand{\tRH}{t_{\text{RH}}}
\newcommand{\tosc}{t_{\text{osc}}}

\definecolor{mypurple}{RGB}{164,64,214}

\newcommand\eea{\end{eqnarray}}
\newcommand\bea{\begin{eqnarray}}

\newcommand{\mpl}{M_{\text{Pl}}}

\newcommand\bes{\begin{split}}
\newcommand\ees{\end{split}}

\newcommand{\eq}[1]{eq.~(\ref{eq:#1})}
\newcommand{\fig}[1]{fig.~\ref{fig:#1}}
\newcommand{\dw}{\ensuremath{\text{dw}}}
\newcommand{\gw}{\ensuremath{\text{gw}}}

\begin{document}

\title{Gravitational waves from the Axiverse}

\date{\today}
\author[a]{Saurav Das,}
\author[a]{Francesc Ferrer}

\affiliation[a]{{Department of Physics and McDonnell Center for the Space Sciences,
Washington University, St. Louis, Missouri 63130, USA}}


\emailAdd{s.das@wustl.edu}
\emailAdd{ferrer@wustl.edu}

\abstract{
Models with axion like particles (ALPs) often predict the formation of a
string-domain wall network in the early universe. We study how such
networks of defects appear in the context of string theory, and discuss the conditions for their long-term stability.
In a scenario with several axions, we show how a bias term in
the potential arises naturally from the effects of multiple instantons,
leading to the eventual decay of the domain walls. 
We find that the annihilation of the network leads to the generation of a
stochastic gravitational wave background (SGWB) with a
spectrum that has characteristic contributions from both walls and strings. The unique shape of the spectrum provides an opportunity to probe
string theory axions at existing and upcoming observatories.
The extinction of the network is also accompanied by the production of different axion mass eigenstates. In a region of the parameter space, 
the lightest eigenstate can be long lived and make up the dark matter in the universe.
}

\maketitle

\section{Introduction}

One of the most general low energy predictions of string theory is the 
presence of multiple axions~\cite{Witten:1984dg, Choi:1985je, Barr:1985hk,Svrcek:2006yi, Arvanitaki:2009fg}. In this context, the four-dimensional
axions are realized as the zero modes of higher dimensional gauge fields. 
A linear combination of these fields can couple to Quantum Chromodynamics
(QCD) in the way required to solve the strong CP 
problem~\cite{Baker:2006ts,Pendlebury:2015lrz,Hook:2018dlk}, 
similar to the Peccei-Quinn (PQ)
axion~\cite{Peccei:1977hh,Peccei:1977ur,Weinberg:1977ma,Wilczek:1977pj}. 
Furthermore, the fact that the masses of string theory axions are generated 
only 
from non-perturbative effects is generically expected to 
ameliorate~\cite{Demirtas:2021gsq} the quality problem of the field theory
PQ axion~\cite{Kamionkowski:1992mf,Barr:1992qq,Ghigna:1992iv,Georgi:1981pu,Lazarides:1985bj,Holman:1992us}. 

In string theory compactifications, the QCD axion is generally accompanied by 
a plethora of other light axionic fields with their own interesting 
dynamics~\cite{Ho:2018qur,Chadha-Day:2021uyt,Foster:2022ajl,Murai:2023xjn}. 
In particular, the axions could be magnetically sourced by string theory 
axion strings that are expected to be present in specific 
models of brane 
inflation~\cite{Dvali:1998pa,Jones:2002cv,Frey:2005jk,Sarangi:2002yt}. Much
like field theory axions, the dynamics of
string theory axions in many axiverse constructions 
is described by a potential generated from 
non-perturbative instanton effects~\cite{Bachlechner:2017zpb,Hu:2020cga}. 
Thus, the $4$D effective field theory (EFT) of the multiple axionic fields 
can then be written as:
\begin{equation}
    \mathcal{L} \supset -\frac{K_{ij}}{2} \partial_\mu a_i \partial^\mu a_j - \Lambda_n^4 \left[ 1 - \cos \left( N_{ni} \frac{a_i}{f_{i}} + \delta_n \right) \right],
	\label{eq:eftinstanton}
\end{equation}
where $K_{ij}$ is the symmetric kinetic mixing matrix, $f_i$ denotes the 
decay constant for the $i$-th axion $a_i$, and summation over repeated indices
is understood. 
$N$ denotes the instanton charge matrix of the axions, and its component
$N_{ni}$ is the charge of the axion $a_i$ under the instanton associated to 
the energy scale $\Lambda_n$. 
We redefine the basis in which $\frac{a_i}{f_i}$ is $2\pi$ periodic, 
i.e. $\frac{a_i}{f_i}= \frac{a_i}{f_i}+2\pi$, so the charges $N_{ni}$ in
\eq{eftinstanton} are all integers. 

Like in the case of the field theory axion, the potential in \eq{eftinstanton} 
may lead to the formation of domain walls (DWs), whose stability depends on 
the instanton charges $N_{ni}$. In the context of field theory, the
analogous question of the stability of DWs has been considered for 
the QCD 
axion~\cite{Sikivie:1982qv, Zeldovich:1974uw,Hiramatsu:2012sc,Hiramatsu:2010yn}
and for the case of DWs of other scalar or pseudoscalar 
fields~\cite{Kibble:1976sj,Larsson:1996sp,Saikawa:2017hiv,Bai:2023cqj,Gouttenoire:2025ofv}. 
In the case of the QCD axion, if the PQ phase transition occurs after inflation~\cite{Sikivie:1982qv,Vilenkin:1982ks,Vilenkin:1984ib,Davis:1986xc,Vincent:1996rb}, a network of global strings is generated. The strings evolve in the
scaling regime until the QCD phase transition, which generates a potential 
for the axion similar to \eq{eftinstanton} with just one axion ($i=j=1$) and 
with instanton charge $N_{\text{DW}}$. 

At this point, DWs appear and attach to
the strings at the location of the maxima of the potential. 
If $N_{\text{DW}}>1$, the axion potential has multiple minima
and the hybrid network of strings and DWs is stable. A 
familiar example is the DFSZ axion model~\cite{Zhitnitsky:1980tq,Dine:1981rt},
with $N_{\text{DW}}= 2 N_{g}$, where $N_g$ is the 
number of generations of fermions that couple to the two Higgs doublets in
the model.
Nevertheless, the long term survival of the DWs is in severe tension with
the observed radiation dominated universe at the time of Big Bang 
Nucleosynthesis (BBN)~\cite{Iocco:2008va}. This is the infamous domain wall 
problem~\cite{Zeldovich:1974uw} that is due to the fact that
the energy density in DWs redshifts 
more slowly than either matter or radiation. Therefore, a DW network 
surviving long enough will eventually come to dominate the energy density of 
the universe~\cite{Sikivie:1982qv}. The universe would then
enter a regime of accelerated expansion that contradicts the observed 
cosmology~\cite{Vilenkin:1984ib}. 

String theory axions are not associated to a PQ phase transitions, and
the formation of topological defects through the Kibble-Zurek
mechanism~\cite{Kibble:1976sj,Zurek:1985qw} does not take place. 
On the other hand, in string theory a network of fundamental cosmic 
strings~\cite{Copeland:2003bj} can be generated by different avenues, such as
D-brane annihilation~\cite{Dvali:1998pa,Jones:2002cv,Sarangi:2002yt}. The
evolution of the resulting axion string network has been considered in
\cite{Cicoli:2022fzy,Benabou:2023npn}, the generation of associated DWs
has received little attention so far. Consider the case when the strings in 
the network source the axion flavor state $a_1$. Due to the potential in
\eq{eftinstanton} we expect the formation of DWs. In this case, the stability 
of the string domain wall network is dictated by the charge $N_{11}$. In 
particular, for $N_{11}=1$ the network is unstable and quickly decays, but
if $N_{11}>1$, the potential allows for multiple degenerate 
minima and the network would be stable. 
In the absence of good priors for the instanton charges, it is usually 
assumed that $N_{ij}$ are arbitrary integers distributed over a large 
range~\cite{Bachlechner:2015qja}. We are, thus, led to conclude that
a stable DW network is a natural possibility in string theory axion models. 

To address the domain wall problem, a mechanism that renders the defect
network unstable is required. In the case of field theory axions, it is
common to introduce a bias term in the potential, which breaks the 
PQ symmetry and removes the degeneracy between different 
vacua~\cite{Sikivie:1982qv,Sikivie:2006ni}. The bias potential can
be generated in UV completions, since the PQ symmetry is expected to be 
violated by Planck suppressed operators~\cite{Harlow:2018tng,Reece:2023czb}. 
As we discuss below, a bias term that leads to the
eventual annihilation of the network naturally appears for string theory 
axions.

The eventual annihilation of the domain walls and the strings has many 
interesting consequences, like the production of a stochastic gravitational 
wave background (SGWB)~\cite{Hiramatsu:2010yz,Kawasaki:2011vv,Hiramatsu:2013qaa,ZambujalFerreira:2021cte,Kitajima:2023cek,Ferreira:2023jbu,Bai:2023cqj,Saikawa:2017hiv} 
that can be observed with 
existing~\cite{KAGRA:2021kbb,NANOGrav:2023gor,Xu:2023wog} and upcoming 
gravitational wave experiments~\cite{Weltman:2018zrl,LISA:2017pwj,Harry:2006fi,Kawamura:2020pcg,Maggiore:2019uih,TianQin:2015yph,Ruan:2018tsw,Reitze:2019iox,Crowder:2005nr,Janssen:2014dka}, or the formation of 
primordial black holes~\cite{Widrow:1989fe,Ferrer:2018uiu,Gelmini:2022nim,Dunsky:2024zdo}. It can also produce mildly boosted axions, which can contribute 
to the observed dark matter density in the 
universe~\cite{Hiramatsu:2010yn,Hiramatsu:2012sc,Kawasaki:2014sqa,Harigaya:2018ooc}. Axion dark matter has been historically explored in the context of QCD 
axions. However, axion-like-particles which are not necessarily connected to 
the strong CP problem are also viable dark matter candidates. For string
theory axions, we will find that the annihilation of DWs can produce more 
than one type of axion dark matter. This provides a complementary observable 
in addition to the stochastic gravitational wave background.


This paper is organized as follows. We delineate the conditions for the 
formation of a string-DW network for string theory axions and its stability
in Section~\ref{sec: domain_wall_formation}; the evolution and subsequent 
decay of the DWs are discussed in Section~\ref{sec: domain_wall_evolution}. 
The phenomenological consequences of the annihilation of the defect network
are explored in Section~\ref{sec: phenomenological_consequences}, and
we conclude in Section~\ref{sec: conclusion}.

\section{Domain walls of string theory axions}
\label{sec: domain_wall_formation}

As mentioned above, in extra dimensional UV completions there is no 
symmetry-breaking PQ phase transition that can seed a network of cosmic
strings by the Kibble-Zurek mechanism\footnote{In~\cite{Cline:2024vbd}
a scenario in which strings are created in a phase
transition associated to the compactification of a warped dimension is
discussed. The transition, however, is first order and not associated to
a PQ symmetry.}. Instead, it is generically expected
that cosmic strings grow around fundamental strings that can form at the
end of D-brane 
inflation~\cite{Dvali:1998pa,Sen:2000vx,Dvali:2002fi,Blanco-Pillado:2005oxi}.
A non-zero field strength remains in the core of the string that can 
magnetically source the axion (see~\cite{Benabou:2023npn} for examples of
this scenario and potential caveats). 
The axion field configuration surrounding an axion string can naturally form 
domain walls. The (in)stability of such string domain wall network depends on 
the details of the axion potential in~\eq{eftinstanton}. If the axion potential
has an single minimum, the domain walls decay shortly after formation. 
Indeed, since in this case each string is connected to one domain wall, the 
tension caused by the domain walls pulls the strings together and they 
quickly annihilate. On the other hand, if the axion potential has multiple 
minima, each string is attached to more than one domain wall, each pulling the
string in different directions. In this case, the network of domain walls is 
stable. For string theoretic axions, the potential in~\eq{eftinstanton}
arises from string theory instantons~\cite{Bachlechner:2017zpb}. In this case,
the number of minima in axion field space is determined by the instanton 
charges of that axion. In the absence of a specific prediction for
the instanton charges, we can take them to be random integers (in the 
normalized basis where axions are $2\pi$ periodic)~\cite{Bachlechner:2015qja}.
Since the magnitude of the charges can naturally be larger than one, the 
formation of a stable string-domain wall network is a natural possibility.

In particular, let us assume that the strings source the axion flavor state 
$a_1$. The instanton potential naturally couples multiple axions, but for the 
sake of simplicity, we specialize to the case of two axions, $a_1$ and $a_2$. 
Among these, $a_1$ is magnetically sourced by the string. The potential 
generated by the instanton associated with the largest scale $\Lambda$ is:
\begin{equation}
    V(a_1,a_2) = \Lambda^4\left[1-\cos\left(N_{1}\frac{a_1}{f_1}+N_{2}\frac{a_2}{f_2}\right)\right],
	\label{eq:potential2axion}
\end{equation}
where we have removed an overall phase through a redefinition of the fields.
Here, $N_1$ and $N_2$ denote the charges of axions $a_1$ and $a_2$ 
respectively. Since the string only sources $a_1$, the stability of the string
domain wall network is dictated by the charge $N_1$ and the domain wall 
number is $N_{\text{DW}}=N_1$. In particular, if 
$N_{1}\neq 1$, the string domain wall network will be stable, the network 
reaches the scaling regime~\cite{Avelino:2005kn,Avelino:2008ve} and quickly 
dominates the energy density of the universe. Such a phase in the cosmic 
history is completely excluded if it occurs close to or after BBN. Although 
the cosmic history before BBN is less constrained, a domain wall dominated 
epoch would cause an inflationary expansion of the 
universe~\cite{Kuster:2008zz} which is difficult to reconcile with the 
observed cosmology. Either way, the string domain wall network needs to 
decay eventually. This decay can occur if the multiple minima are not 
completely degenerate. The energy bias then causes a pressure on different
sides of the DWs that leads to their annihilation. For concreteness, we 
assume that a different instanton generated potential lifts the 
degeneracy: 
\begin{equation}
    V_b(a_1) =\frac{\Lambda_b^4}{2}
	\left[1- \cos\left(N_b\frac{a_1}{f_1}\right) \right]
\end{equation}
where $\Lambda_b \ll \Lambda$ denotes the scale of the instanton that is 
responsible for the bias potential, and $N_b$ the charge of the axion $a_1$ 
under the same instanton. We elaborate on the general condition for an 
instanton generated potential to act as a bias potential in 
Appendix~\ref{App: charge_condition}. As a simple example, we 
assume $N_2=1$ and $N_b=1$. This particular choice of charges removes the 
degeneracy between multiple minima and leads to the collapse of the network.  
We also assume the all the string instanton generated potentials are 
independent of the temperature. It should be noted that, since the energy 
density of the DWs and other relevant quantities depend linearly on 
$N_{\text{DW}}$, a different choice of the charges would change
the numerical details of the phenomenological signatures by an
$\mathcal{O}(1)$ factor.


\section{Evolution and decay of the string-domain wall network}
\label{sec: domain_wall_evolution}
The thickness $\delta$ of the domain walls is governed by the mass of $a_1$, 
which is generated by the leading potential $V$:
\begin{equation}\label{Eq: t_delta}
    \delta^{-1} = \frac{N_1\Lambda^2}{f_1}.
\end{equation}
The domain walls are effectively ``produced'' when the horizon size becomes 
smaller than the thickness of the domain walls, $H\sim \delta^{-1}$. 
At this time, the energy density of the network is dominated by the strings. 
In the case of field theory axions, the energy density of the string 
component of the network in the scaling 
regime~\cite{Hiramatsu:2012gg,Hiramatsu:2012sc,Yamaguchi:1998gx,Yamaguchi:2002sh,Yamaguchi:2002zv} is given by:
\begin{equation}
\label{Eq: QCD_axion_string_energy_density}
    \rho_s(t) =\xi\frac{\mu}{t^2},
\end{equation}
where $\xi$ is an $\mathcal{O}(1)$ number, $\mu$ is the tension in the strings
and $t$ is time. String theory axion strings are qualitatively different from
field theory axion strings~\cite{March-Russell:2021zfq,Dolan:2017vmn,Reece:2018zvv,Lanza:2021udy,Heidenreich:2021yda,Benabou:2023npn}. The structure of the 
strings depends on the details of the compactification of the extra 
dimensions, but if warping is not present, the string tension is given 
by~\cite{Benabou:2023npn}:
\begin{equation}
\label{Eq: string_tension}
    \mu = \kappa f_1 \mpl,
\end{equation}
where $\kappa$ is an $\mathcal{O}(1)$ number and 
$\mpl=\frac{1}{\sqrt{8\pi G}}$ denotes the reduced
Planck mass. This is in contrast with field theory axion 
strings\footnote{If warped extra dimensions are involved, like in the 
scenario presented in~\cite{Cline:2024vbd}, the tension could be similar to 
that of field theory axions.}, which have 
a parametrically lower tension $\mu\propto f_1^2$.

The total energy density in the string theory axion strings is determined by 
the large number of subhorizon-size string loops. The differential number 
density of string loops of length $l$ per unit volume is given by $n(l,t)$, 
so that the total number of strings in a unit volume is $n(t)=\int n(l,t)
\dd{l}$. The simulations 
differ widely on the size and number density of 
loops~\cite{Blanco-Pillado:2013qja,Ringeval:2005kr,Lorenz:2010sm}, and
for concreteness, we use the prescription 
in~\cite{Blanco-Pillado:2013qja,Benabou:2023npn}. 
The differential number density in the scaling regime is: 
\begin{align}
    n(l,t)\approx 
	\frac{\alpha t^{-4}}{\left(\frac{l}{t}+r_G G\mu\right)^{5/2}},
\end{align}
where $\alpha\approx 0.18$ and $r_G\approx \mathcal{O}(50)$ are model 
dependent constants~\cite{PhysRevD.31.3052,Allen:1991bk,Allen:1994iq,Blanco-Pillado:2013qja}, and $G$ is the gravitational constant. 
The expected string length $\langle l(t)\rangle$ is related to the size of the
horizon by:
\begin{align}
    \langle l(t)\rangle \approx H^{-1}\frac{f_1}{\mpl}.
\end{align}
We, thus, find that the total energy density in the strings is:
\begin{align}
    \rho_s(t) \approx \mu \langle l(t)\rangle \langle n(t)\rangle
	\approx \frac{\mu}{t^2}\sqrt{\frac{\mpl}{f_1}}.
\end{align}
This is parametrically larger than the energy density of field theory axion 
strings given in eq.~\ref{Eq: QCD_axion_string_energy_density}, which
has a number of important phenomenological consequences. 
The contribution of the strings to the total energy density of the 
string-domain wall network remains important until much later compared to the 
standard field theory PQ string. Even more importantly, the GW signal
and the amount of axionic DM produced by the strings are much larger. 

As for the domain walls, their energy density is negligible compared to the 
strings immediately after they enter the horizon. For a stable string-domain 
wall network, the walls themselves quickly reach a scaling 
regime~\cite{Martins:2016ois}. As long as the DWs do not dominate the energy 
density of the universe, their energy density in the scaling regime
behaves as:
\begin{equation}
\label{Eq: energydensityDW}
    \rho_{\dw}(t)=\mathcal{A}\frac{\sigma}{t},
\end{equation}
where the so-called ``area parameter'' is roughly 
given by $\mathcal{A} = (0.8\pm 0.1)\NDW$, with $\NDW = N_1$ in our case,
and the surface mass density of the domain walls is 
$\sigma = 8f_1\Lambda^2$ ~\cite{Hiramatsu:2012sc}. 

Since the energy density of the strings redshifts faster than that of the 
domain walls, after a time 
$t_*\sim \frac{\mpl}{\Lambda^2}\sqrt{\frac{\mpl}{f_1}}$, the strings become
subdominant. The temperature $T_*\equiv T(t_*)$ of the universe at this time
is:
\begin{equation}
\label{Eq: T_star}
    T_*\simeq \Lambda\left(\frac{f_1}{\mpl}\right)^\frac{1}{4}
	\left(\frac{45}{2\pi^2 g_*}\right)^\frac{1}{4}
\end{equation}
where $g_*\equiv g_*(T_*)$ is the effective number of degrees of freedom,
and we have neglected factors of order unity like $\mathcal{A}$ 
in this expression for simplicity. 

From this point onward, the domain walls dominate the energy density of the 
universe unless they decay. The eventual demise of the domain walls
is brought about by the bias potential, which produces a difference in the 
potential energy between two successive local minima:
\begin{equation}
    \Delta V \approx \Lambda_b^4\sin^2\left(\frac{\pi N_b }{N_1}\right)
	= \Lambda_b^4
\end{equation}
where in the last equality we have used the values $N_1=2$ and $N_b=1$. 
The difference in potential energy acts as a pressure against the domain walls
and accelerates them towards the true vacuum. When the pressure force roughly equals the surface tension of the walls, the domain walls decay. This occurs
at the annihilation time:
\begin{equation}\label{Eq: tann}
    t_{\text{ann}}= C_{\text{ann}}\frac{\mathcal{A}\sigma}{\Delta V}
	\approx C_{\text{ann}}\frac{8\mathcal{A}f_1\Lambda^2}{\Lambda_b^4},
\end{equation}
where the proportionality constant $C_{\text{ann}}$ can be obtained from 
numerical simulations~\cite{Hiramatsu:2012sc}. In the case of domain walls
associated to a $Z_N$ symmetry~\cite{Hiramatsu:2013qaa}, 
$C_{\text{ann}}\approx 10-20$, independent of the domain wall number. 
Following~\cite{Hiramatsu:2013qaa}, we take $C_{\text{ann}}=18$ as a benchmark
value for the remainder of the paper.

Assuming that the universe is radiation dominated throughout the evolution of 
the string-domain wall network, we can estimate the temperature of the thermal
plasma at the time of annihilation:
\begin{equation}
    T_{\text{ann}} \equiv T(t_{\text{ann}}) \approx 
	\frac{\Lambda_b^2}{\Lambda}
	\left(\frac{\mpl}{f_1}\right)^{\frac{1}{4}},
\end{equation}
where we have again ignored $\mathcal{O}(1)$ factors for clarity. The 
fractional energy density of the domain walls at the time of annihilation is 
given by:
\begin{equation}
	\Omega_{\text{DW}}(t_{\text{ann}}) = 
	\frac{4\cdot 8^2}{3} \mathcal{A}^2 
	C_{\text{ann}} \frac{ f_1^2 \Lambda^4}{\Lambda_b^4 \mpl^2},
	\label{eq:omega_wall}
\end{equation}
where $\Omega_{\text{DW}}=\frac{\rho_{\text{DW}}}{\rho_c}$ is the 
ratio of the DW energy density to the critical energy density. 

\begin{figure}[t]
    \centering
    \includegraphics[width=0.75\textwidth]{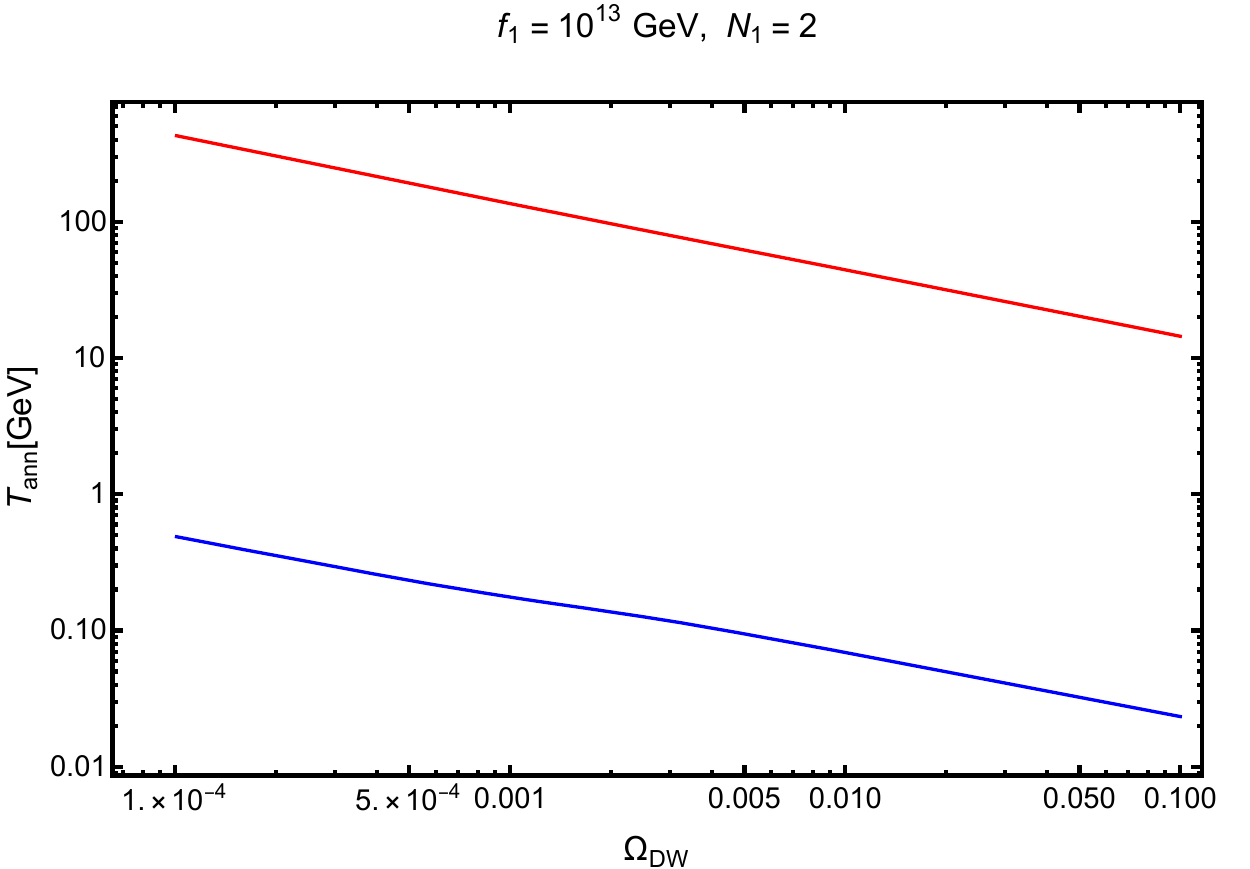}
    \caption{ The annihilation temperature of the domain walls as a function of the
	fractional energy density of the domain walls assuming a radiation dominated 
	universe. The red (blue) denotes $\Lambda=1~\text{TeV}~(\Lambda=1~\text{GeV})$. We 
	see that for natural values of the parameters $\Lambda$ and $f$, the annihilation 
	can happen during electroweak phase transition or at the time
	of the QCD phase transition.}
    \label{fig: Omega_DW}
\end{figure}

For a string theory axion model with a given scale $\Lambda$ in 
\eq{potential2axion}, we expect that decreasing the magnitude of the bias 
potential will push the annihilation of the 
network to lower temperatures with the DWs attaining a larger energy density. 
This is borne out in \fig{ Omega_DW}, which shows that for reasonable values 
of the parameters the annihilation can take place at the time of the QCD 
phase transition, but it could
also occur earlier in the evolution of the universe, e.g. during the 
electroweak phase transition.

\subsection{Size of the bias term}

A domain wall dominated era produces a period of late inflationary expansion,
which is extremely difficult to make it compatible with the observed 
cosmology~\cite{Kuster:2008zz}. 
The size of the bias potential to avoid a domain wall dominated era can be 
calculated as follows. As usual, we consider the domain wall evolution in a 
radiation dominated era, and we defer the discussion of the case of an
early matter dominated era (EMD) to app.~\ref{App:emd}. The Hubble parameter 
at the time of annihilation is given by
\begin{equation}
    \begin{split}
        H_{\text{ann}} = 
	    \frac{1}{2}\frac{\Delta V}{C_{\text{ann}}\mathcal{A}\sigma}.
    \end{split}
\end{equation}
Demanding that at the time of annihilation, the total energy of the universe 
is larger than the energy density in the domain walls, 
i.e. $\rho_{\text{tot}}>\rho_{\text{DW}}$, we obtain the constraint:
\begin{equation}\label{Eq: bias_minimum}
    V_b\geq \frac{32\pi }{3}\mathcal{A}^2C_{\text{ann}}\sigma^2G
	\implies
	\Lambda_b\geq\left(\frac{4\pi}{3}
	\mathcal{A}^2C_{\text{ann}}\right)^{\frac{1}{4}}
	\Lambda\sqrt{\frac{8f_1}{\mpl}}.
\end{equation}

In addition to producing a pressure force on the network, the presence of the 
bias potential alters the probability of the axion field choosing
the different vacua (which are no longer degenerate). This unequal population 
of the different vacua in turn biases the domain walls and can also 
cause the network to 
decay~\cite{Coulson:1995nv,Larsson:1996sp,Hiramatsu:2010yn}. But, as shown 
in app.~\ref{App : Unequal_abundance}, the timescale of annihilation due to 
this effect is much larger than the timescale corresponding to the pressure 
force. 

On the other hand, a large bias potential can cause the collapse of the 
string domain-wall network to take place much earlier.
In particular, if the bias potential is sufficiently large, the domain wall
annihilation time given in eq.~\ref{Eq: tann} can occur before $t_*$. 
In that case, the strings dominate the energy density of the network at the 
time of annihilation. This is avoided if the bias potential remains below:
\begin{equation}\label{Eq: bias_maximum}
    \Lambda_b \leq \Lambda\left(\frac{f_1}{\mpl}\right)^{\frac{3}{8}}.
\end{equation}

\section{Phenomenological signatures}
\label{sec: phenomenological_consequences}
As soon as the condition for the decay of the domain walls is met, the network
quickly annihilates at the time given in eq.~\ref{Eq: tann}. More 
specifically, the fraction of the original network that survives at time 
$t>t_{\text{ann}}$ can be approximated as:
\begin{equation}
\label{Eq: survival_prob}
	f_{\text{DW}} \sim
	\exp\left[-\left(\frac{t}{t_{\text{ann}}}\right)^{\alpha}\right].
\end{equation}
Although the small number of surviving domain walls makes it challenging to 
properly estimate this probability, numerical simulations suggest 
$\alpha = 3/2$~\cite{Ferreira:2024eru}. This small fraction of late collapsing walls can contribute significantly to the production of primordial black
holes~\cite{Ferrer:2018uiu}. We will not further discuss this interesting
signature in quantitative detail given the uncertainties 
involved~\cite{Dunsky:2024zdo}. In the following, we focus instead on
the generation of gravitational waves and the production of axion dark
matter, which mainly take place within one Hubble time of $t_{\text{ann}}$.

\subsection{Production of Gravitational waves}

Gravitational waves provide a handle to high energy phenomena in the early universe. As we
show below and summarize in \fig{ perturbation}, the different contributions to the SGWB from
the DWs and from the strings generate a unique spectral shape that could allow present
and forthcoming experiments to unveil string theory axions.

\subsubsection{Gravitational waves from domain walls}

The production of gravitational waves from the decay of domain walls depends 
on whether they decay during a radiation dominated epoch or during the more 
exotic possibility of an early matter dominated epoch. We describe here the
former scenario, and we relegate the study of the EMD case to 
app.~\ref{App:emd}. In appendix~\ref{App: No_EMD}, we show that the axions 
produced from the annihilation of strings in the scaling regime can not 
lead to a matter dominated era until the collapse of the whole network. 
Hence, our assumption that network evolves and collapses in the radiation 
dominated era is self consistent.

The power radiated in GWs from the annihilation of domain walls can be 
estimated from the leading quadrupole 
approximation~\cite{10.1093/acprof:oso/9780198570745.001.0001}:
\begin{equation}
    P \sim G \dddot{Q}_{ij}\dddot{Q}_{ij}.
	\label{eq:quadrupole}
\end{equation}
The quadrupole moment of a domain wass can be estimated as 
$Q\sim M_{\text{DW}}L(t)^2$, where $M_{\text{DW}}\sim \sigma L(t)^2$ is the 
mass of the wall and $L(t)$ is its curvature radius, which is $\propto t$ in 
the scaling regime. Putting these expressions in \eq{quadrupole}, we get 
$P\sim  G \sigma^2 t^2$. The energy density emitted in GWs is:
\begin{equation}
    \rho_{\gw}\sim E_{\gw}/t^3\sim G \mathcal{A}^2\sigma^2.
\end{equation}
The power spectrum of the GWs from DW annihilation is: 
\begin{equation}
\label{Eq: GW_power_spectrum_definition}
	\Omega_{\gw}(k,t)=\frac{1}{\rho_c(t)} \dv{\rho_{\gw}(k,t)}{\log k}.
\end{equation}
The spectrum has a peak in wavenumber corresponding to the annihilation 
timescale $k_{\text{peak}}(t_{\text{ann}})\sim H(t_{\text{ann}})$. The 
amplitude of the spectrum is given by:
\begin{equation}
\label{Eq: GWfromDW}
	\pqty{\dv{\rho_{\gw}}{\log k}}_{\text{peak}}= 
	\epsilon_{\gw} G \mathcal{A}^2\sigma^2,
\end{equation}
where $\epsilon_{\gw}$ is an $\mathcal{O}(1)$ proportionality factor that can 
be inferred from numerical simulations~\cite{Hiramatsu:2012sc}. For example,
$\epsilon_{\gw}=5-10$ in the case of
domain walls associated to a complex scalar with a $U(1)$ symmetry
broken to a $Z_N$~\cite{Hiramatsu:2012sc}. However, it should be noted that 
there are significant uncertainties in $\epsilon_{\gw}$. For instance, as 
found in dedicated numerical simulations of the production of GWs in a 
radiation dominated universe~\cite{Ferreira:2024eru,Kitajima:2023cek}, the 
domain walls ``effectively'' decay at $t\approx 3 t_{\text{ann}}$, which 
increases the amplitude of the GW power spectrum by a factor of $\sim 10$. 
Nevertheless, to obtain a conservative estimate of the amplitude of the GW 
background, we will assume that the network annihilates at 
$t\approx t_{\text{ann}}$. 

The frequency scale corresponding to the highest power, $f_{\text{peak}}$, is 
simply related to the Hubble parameter during annihilation:
\begin{equation}
	f_{\text{peak}} = H(t_{\text{ann}})\frac{R(t_{\text{ann}})}{R(t_0)}
	= H(t_{\text{ann}})\frac{T_0}{T_{\text{ann}}},
\end{equation}
where $R(t)$ denotes the scale factor at time $t$, and the subscript $0$ 
refers to the present epoch. In the last equality we have assumed a radiation 
dominated universe. Since $H(t_{\text{ann}})\propto T_{\text{ann}}^2$, the 
spectrum peaks at higher frequencies as the annihilation happens earlier. 
For lower frequencies, $f<f_{\text{peak}}$, the spectrum has the usual causal 
tail~\cite{Caprini:2018mtu,Cai:2019cdl} $\Omega_{\gw}\propto f^3$; and the 
power decays as $\Omega_{\gw}\propto f^{-1}$ for higher frequencies, 
$f_{\delta}>f>f_{\text{peak}}$, as confirmed by numerical 
simulations~\cite{Hiramatsu:2012sc}. The frequency $f_{\delta}$ is another 
important scale in the spectrum, and it corresponds to the width of the 
domain walls: 
\begin{equation}
    f_{\delta} = \delta^{-1}\frac{R(t_{\text{ann}})}{R(t_0)}= \delta^{-1}\frac{T_{\text{ann}}}{T_0}.
\end{equation}
Since this is the smallest scale in the network, the power at scales below this
one is extremely suppressed, as noted in the 
simulations~\cite{Hiramatsu:2012sc}. The precise scaling of the power spectrum
beyond this frequency is difficult to obtain from numerical simulations, 
although there are some hints that the power increases as the number of domain
walls, $N_{DW}$, increases. The GW energy density at the time of annihilation
is given by:
\begin{equation}
\label{Eq:GW_amplitude}
    \Omega_{\gw}(t_{\text{ann}})_{\text{peak}} = 
	\frac{1}{\rho_c \pqty{t_{\text{ann}}}} 
	\pqty{\dv{\rho_{\text{gw}}(t_{\text{ann}})}{\log k}}_{\text{peak}} 
	= \frac{8\pi \epsilon_{\gw} G^2 \mathcal{A}^2 \sigma^2}{3 
	H^2(t_{\text{ann}})}.
\end{equation}
As noted above, the majority of the domain walls decay shortly after 
$t_{\text{ann}}$. In the limit of instantaneous decay, the GW power spectrum 
measured today is given by:
\begin{equation}
\begin{split}
    \Omega_{\gw} h^2 (t_0) &= \frac{\rho_{\gw}(t_0) h^2}{\rho_c(t_0)} = 
	\frac{\rho_c(t_{\text{ann}}) h^2}{\rho_c(t_0)} \left( 
	\frac{R(t_{\text{ann}})}{R(t_0)} \right)^4 \Omega_{\gw}
	(t_{\text{ann}})\\
	&= \Omega_{\text{rad}} h^2 \left( \frac{g_*(T_{\text{ann}})}{g_{*0}} 
	\right) \left( \frac{g_{*s0}}{g_{*s}(T_{\text{ann}})} \right)^{4/3} 
	\Omega_{\gw} (t_{\text{ann}}),
\end{split}
\end{equation}
where $g_{*0}= 3.36$ and $g_{*s0}= 3.91$ are the number of relativistic 
degrees of freedom for the energy density and the entropy density respectively
at the present time; while $g_*(T_{\text{ann}})$ and $g_{*s}(T_{\text{ann}})$ 
are the corresponding values at $t=\tann$. We have used the formulae 
in~\cite{Saikawa:2018rcs} to numerically evaluate $g_*(T)$ and $g_{*s}(T)$.
\begin{figure}[t]
    \centering
    \includegraphics[width=0.75\textwidth]{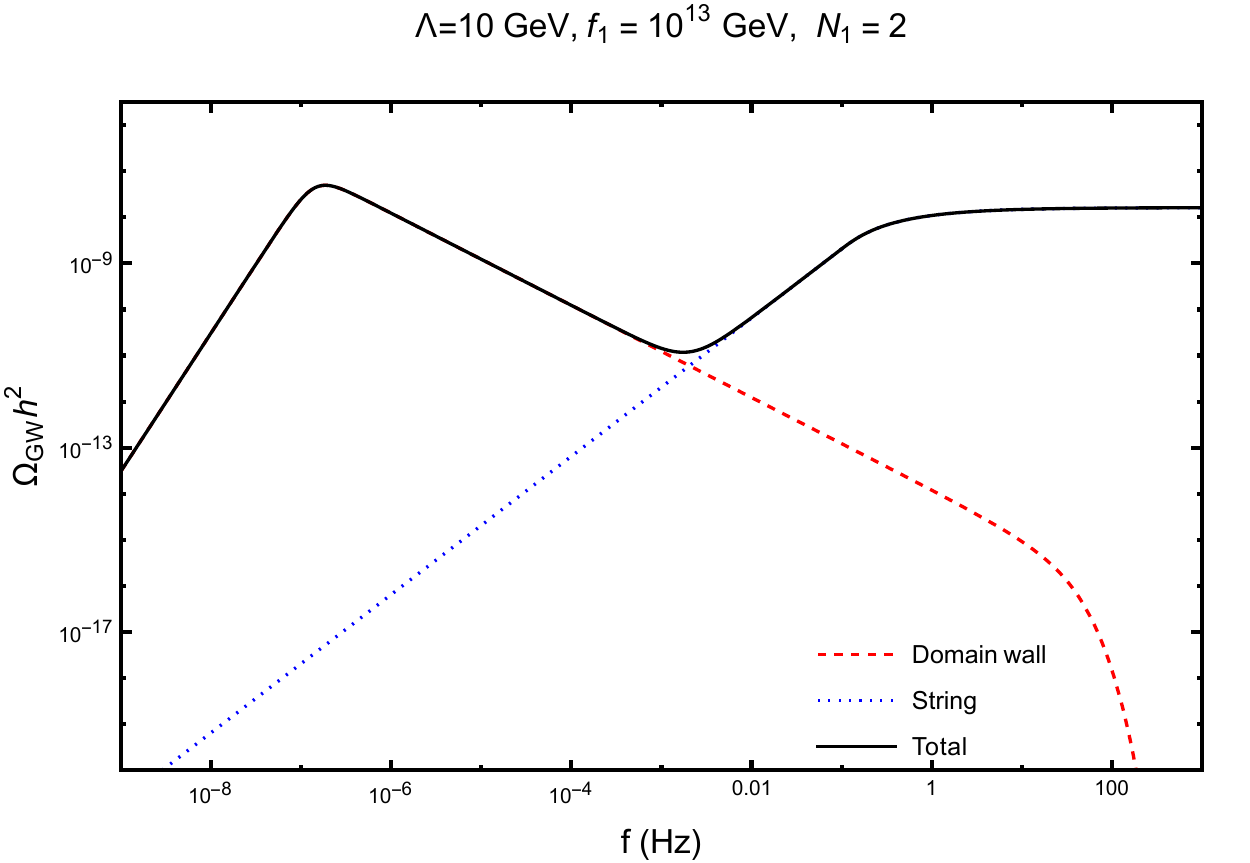}
	\caption{The red-dashed curve shows the gravitational wave spectrum 
	produced by the domain walls for $\Lambda=10$ GeV, $f_1=10^{13}$
	GeV, $N_1=2$ and $\rho_{\dw}(t_{\text{ann}}) = 
	10^{-2}\rho_{\text{tot}}(t_{\text{ann}})$. The blue-dotted line
	shows the spectrum produced by the string network, and the combined 
	spectrum is depicted by the black curve. String theory axion 
	strings produce larger amounts of GWs because of their higher tension.
	The high energy tail corresponds to the logarithmic divergence in the 
	total GW energy emitted by the strings.}
    	\label{fig: GW_DW+string}
\end{figure}

\subsubsection{Gravitational waves from strings in the scaling regime}

Until now, we have not considered the energy density of the strings in 
discussing the GW power spectrum. However, depending on the size of the bias 
potential, the network might annihilate at a time when the strings dominate 
the net energy density of the network. In particular, when the bias potential 
is sufficiently large to violate eq.~\ref{Eq: bias_maximum}, the contribution
of the strings to the GW power spectrum can no longer be neglected. 

The energy density of the strings in the scaling regime redshifts as 
$\rho_s\sim \frac{\mu}{t^2}\sqrt{\frac{\mpl}{f_1}}$. Conservation of energy 
requires the strings to continuously radiate energy at a rate $\Gamma\sim 
\dd{\rho_s}/\dd{t}$, where $\Gamma$ denotes the energy radiated by the string 
network per unit time per unit volume. The strings emit gravitational 
waves and axions, but under certain assumptions~\cite{Benabou:2023npn} the
respective emission rates are parametrically equal. 
The energy radiated in GWs per unit time per unit volume from string loops 
is:
\begin{equation}
    \Gamma_{\gw}=\frac{16\alpha \, r_G\sqrt{8\pi\kappa}}{3r^{3/2}} 
	H^3\sqrt{f_1\mpl^3},
\end{equation}
where we have used eq.~\ref{Eq: string_tension} for the string tension $\mu$;
the constants $\kappa$, $\alpha$, $r_G$ were defined in 
sec.~\ref{sec: domain_wall_evolution}; and $r=r_G+r_a\frac{8\pi}{\kappa^2}$ 
accounts for GW as well as axion emission, with 
$r_a\approx \mathcal{O}(10)$~\cite{Davis:1985pt,Vilenkin:1986ku} a 
dimensionless constant. 

The GW spectrum is determined by how the total power is distributed among 
gravitational radiation at different frequencies. The spectrum of 
gravitational radiation emitted by a single loop of length $\ell$ depends on 
the shape of the loop, with a lower cutoff frequency $\omega_\ell \geq 4\pi/
\ell$. For an ensemble of string loops, the spectrum can be approximated by a 
power law. Namely, the energy emitted from loops of length $\ell$ is: 

\begin{align}
	\dv{\dot{E}^\ell_\text{GW}}{\omega} =r_G G\mu^2
    \begin{cases} 
    0 & \omega < \omega_\ell \\
    \omega_1^{q-1} \frac{q-1}{\omega^q} & \omega > \omega_\ell .
\end{cases}
\end{align}
The total GW emission per unit time per unit volume averaged over the string 
network is then:
\begin{align}
	\dv{\Gamma_{\gw}(t,\omega)}{\omega}=\int{ \dd{\ell} n(\ell,t)
	\dv{\dot{E}^\ell_\gw}{\omega}}.
\end{align}
Finally, the total energy density in GWs at some time $t$ can be calculated 
from the emission rate: 

\begin{align}
\label{Eq: rho_GW_omega_definition}
	\dv{\rho_{\gw}(t,\omega)}{\omega}=\int_0^t{\dd{t'}
	\left(\frac{R(t')}{R(t)}\right)^3 
	\dv{\Gamma_{\gw}(t',\omega')}{\omega'}},
\end{align}
where $\omega'=\omega R(t)/R(t')$ accounts for the redshift of the GW 
frequency. To obtain the gravitational wave energy density emitted by the 
string network, one needs to numerically evaluate 
eq.~\ref{Eq: rho_GW_omega_definition} for a specific cosmological history. 
However, under the assumption that the number of radiation-like degrees of 
freedom remains unchanged throughout the evolution of the strings,
it is possible to write the total gravitational wave energy density as:
\begin{equation}
\label{Eq: rho_GW_omega}
    \dv{\rho_{\gw}(\omega, t)}{\omega} = 
	\Gamma_{\gw} \frac{x_{\mathrm{UV}}}{5 \cdot 8 \pi \cdot H^2}
	\mathcal{F} \pqty{\omega/\omega_0},
\end{equation}
where $x_{\mathrm{UV}}=r_GG\mu$, $\omega_0 = 8\pi H/x_{\mathrm{UV}}$,
and we have defined:
\begin{equation}
	\label{eq:fx}
	\mathcal{F}(x) \equiv 
	\begin{cases}
		\frac{4(q-1)}{2q+1} \sqrt{x}\,, &\quad\text{if } x \leq 1 \\[10pt]
		\frac{1}{2-q}\left[3(q-1)x^{-2}
		-\frac{15}{2q+1} x^{-q} + 5(2-q)x^{-1}\right]\,, &\quad
		\text{if }x > 1.
	\end{cases}
\end{equation}
The string network produces gravitational radiation until the annihilation of 
the network at time $\tann$. Afterwards, the gravitational waves redshift 
without significant interaction with the matter and radiation. The 
gravitational wave power spectrum observed today is then:
\begin{align}
\label{Eq: GW_power_spectrum}
	h^2 \dv{\Omega_{\gw,0}}{\log\omega} &\equiv 
	\frac{h^2}{\rho_{c,0}} \dv{\rho_g (t_0, \omega)}{\log \omega}
	\nonumber \\
	&= \frac{h^2}{\rho_{c,0}} \left(\frac{R_{\text{ann}}}{R_0}\right)^4 
	\dv{\rho_g(t_{\text{ann}}, \omega R_0/R_{\text{ann}})}{\log \omega}
	\nonumber \\
	&= \frac{16 \sqrt{2\pi^5}}{2025} \frac{g_{*,\text{ann}} 
	g_{*,s,0}^{4/3}}{g_{*,s,\text{ann}}^{4/3}} 
	\frac{h^2 T_0^4}{H_0^2 \mpl^2} 
	\left(\frac{r_G \sqrt{\kappa \alpha}}{r^{3/2}}\right) 
	\sqrt{\frac{f_{1}}{\mpl}} \eval{x\mathcal{F}(x)}_{x = 
	\frac{R_0 \omega}{R_{\text{ann}} \omega_{0,\text{ann}}}}.
\end{align}
Fig.~\ref{fig: GW_DW+string} shows the stochastic gravitational 
wave background produced by the string network, assuming that the emission is 
dominated by kinks, $q=5/3$~\cite{Auclair:2019wcv}. Interestingly,
the total GW spectrum shows the distinctive signature of the low frequency 
peak from the domain wall decay, as well as the high frequency tail from the 
strings. Observation of such a singular GW background would be a smoking gun
of string theory axion strings. For reasonable values of the parameters,
\fig{ perturbation} shows how
this can be achieved through a combination of low-frequency radio telescope
observations that could probe the DW contribution, together with observations
at higher frequencies that could map the contribution from strings
(e.g. by space-borne LISA or next-generation ground based GW observatories).

\begin{figure}[t]
    \centering
    \includegraphics[width=0.75\textwidth]{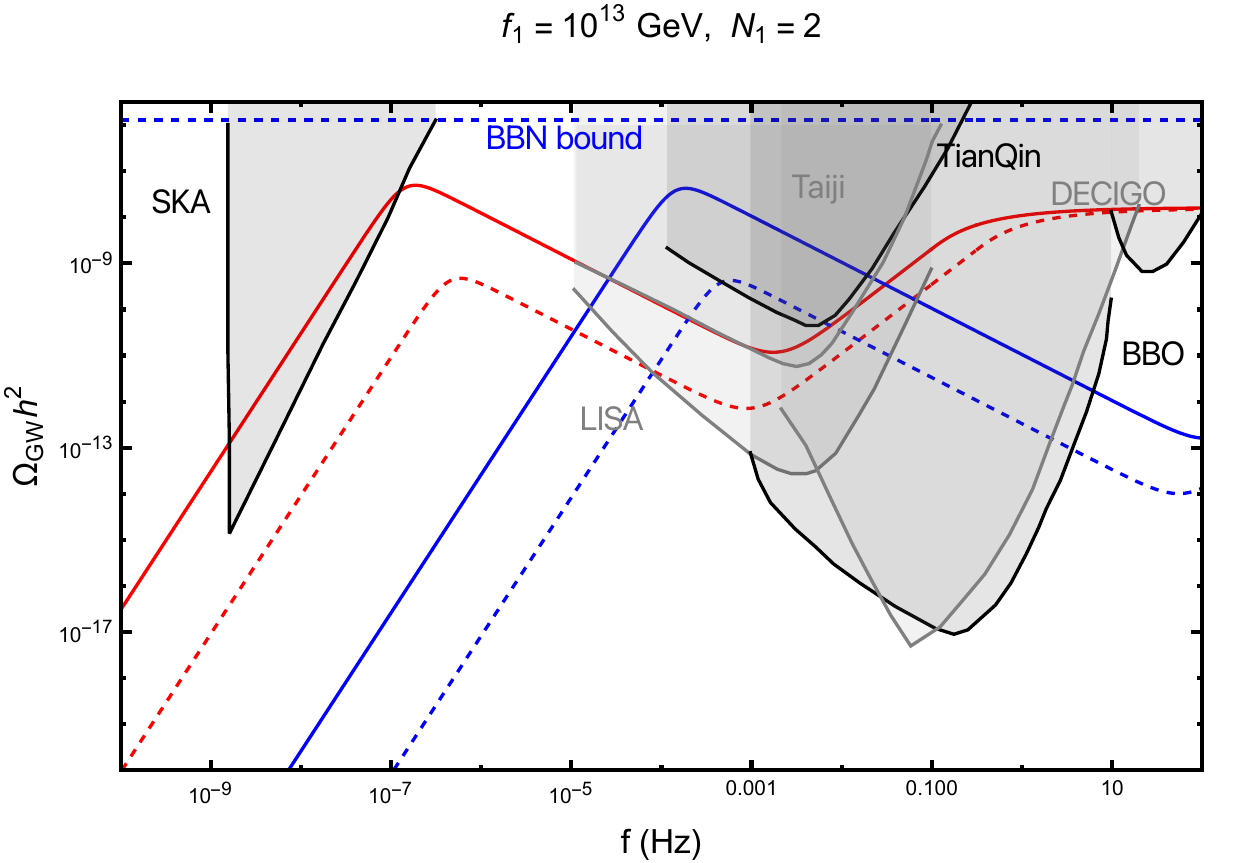}
    \caption{The gravitational wave spectrum produced by the collapse of 
	domain walls with instanton scales $\Lambda = 10~\text{GeV}$ (red) 
	and $\Lambda = 10~\text{TeV}$ (blue). The axion decay constant and the
	domain wall number are set 
	to $f_1=10^{13}~\text{GeV}$ and $N_1 = 2$. The solid and the dashed 
	lines depict $\rho_{\text{DW}}(t_{\text{ann}}) = 10^{-2}\rho_{\text{tot}}(t_{\text{ann}})$ and $\rho_{\text{DW}}(t_{\text{ann}}) = 10^{-3}\rho_{\text{tot}}(t_{\text{ann}}) $ respectively.}
    \label{fig: perturbation}
\end{figure}

\subsection{Production of axion dark matter}\label{subsection: axionic_DM}

The annihilation of the network produces mildly relativistic 
axions~\cite{Hiramatsu:2012sc}. The network emits the flavor eigenstate $a_1$ 
it couples to, but the abundance of axions in the late universe is dictated 
by the properties of the mass eigenstates. As noted in~\cite{Benabou:2023npn},
kinetic mixing does not alter the abundance of axion mass eigenstates barring 
some extreme fine-tuned scenarios where the mass mixing is canceled exactly 
due to kinetic mixing. Therefore, we can ignore kinetic mixing without 
forgoing any interesting property of the axion mass eigenstates. 

The mass matrix for the axion flavor states is given by (using the notation 
in~\cite{Benabou:2023npn}):
\begin{equation}
\label{Eq: mass_matrix}
    \mathcal{L} \supset -\frac{1}{2}
\begin{pmatrix}
a_1 \\
a_2
\end{pmatrix}^\top
\begin{pmatrix}
M^2_{a_1} & M^2_{a_1 a_2} \\
M^2_{a_1 a_2} & M^2_{a_2}
\end{pmatrix}
\begin{pmatrix}
a_1 \\
a_2
\end{pmatrix},
\end{equation}
where the elements of the mass matrix are:
\begin{equation}
    \begin{split}
        &M^2_{a_1}= \frac{N_1^2\Lambda^4+N_b^2\Lambda_b^4}{f_{1}^2},\\
        &M^2_{a_2}= \frac{N_{2}^2\Lambda^4}{f_{2}^2},\\
        & M^2_{a_1a_2}=\frac{N_{1}N_{2}\Lambda^4}{f_{1}f_{2}}.
    \end{split}
\end{equation}
The mass eigenstates $(\phi_2,\phi_1)$ obtained after diagonalizing 
eq.~\ref{Eq: mass_matrix} can be written in terms of rotated flavor states 
$(a_1,a_2)$ as:
\begin{equation}
    \begin{pmatrix}
        \phi_2 \\
        \phi_1
    \end{pmatrix} = R(\theta)\begin{pmatrix}
        a_1 \\
        a_2
    \end{pmatrix},
\end{equation}
where
\begin{equation}
    R(\theta) = \begin{pmatrix}
        \cos \theta & \sin \theta \\
        -\sin\theta & \cos\theta
    \end{pmatrix}.
\end{equation}
The rotation angle is: 
\begin{equation}
\label{Eq: rotation_angle}
    \begin{split}
        &\cos 2\theta = \frac{M^2_{a_1}-M^2_{a_2}}{\Delta^2},\\
        & \sin 2 \theta = \frac{2M^2_{a_1a_2}}{\Delta^2},
        \\
        & (\Delta^2)^2 = (M^2_{a_1}-M^2_{a_2})^2+4 (M^2_{a_1a_2})^2.
    \end{split}
\end{equation}
And we finally obtain the masses of the mass eigenstates:
\begin{equation}\label{Eq: L1Lb}
\begin{split}
    M_{\varphi_1}^2 &= \frac{1}{2} \left( M_{a_1}^2 + M_{a_2}^2 - 
	\Delta^2 \right), \\
    M_{\varphi_2}^2 &= \frac{1}{2} \left( M_{a_1}^2 + M_{a_2}^2 + 
	\Delta^2 \right),
\end{split}
\end{equation}
where we have assumed that $\phi_2$ is the heavier mass eigenstate, 
$M_{\phi_2}>M_{\phi_1}$. 

Some important properties of the mass eigenstates are immediately apparent. 
The constraint that must be satisfied between the instanton scales $\Lambda$ 
and $\Lambda_b$ to prevent domain walls from dominating the energy density
of the universe is given by eq.~\ref{Eq: bias_minimum}, which can be 
schematically cast in the form $\Lambda_b\sim\Lambda\sqrt{\frac{f_1}{\mpl}}$. 
From \eq{omega_wall}, we deduce that the scale of the bias potential 
that corresponds to the domain walls having a given fractional energy density 
is given by $\Lambda_b\sim \Omega_{DW}(t_{\text{ann}})^{-\frac{1}{4}}\Lambda\sqrt{\frac{f_1}{\mpl}}$. 
Hence, in order for the the domain walls to have any observable effects, we 
need the instanton scales to be well separated, $\Lambda \gg \Lambda_b$. This 
suggests that the heavier mass eigenstate is predominantly the linear 
combination of flavor eigenstates that couples to the instanton with the 
larger scale, $\phi_2\sim N_{1}a_1+N_{2}\left(\frac{f_1}{f_2}\right)a_2$.

\begin{figure}[t]
    \centering
    \includegraphics[width=0.85\textwidth]{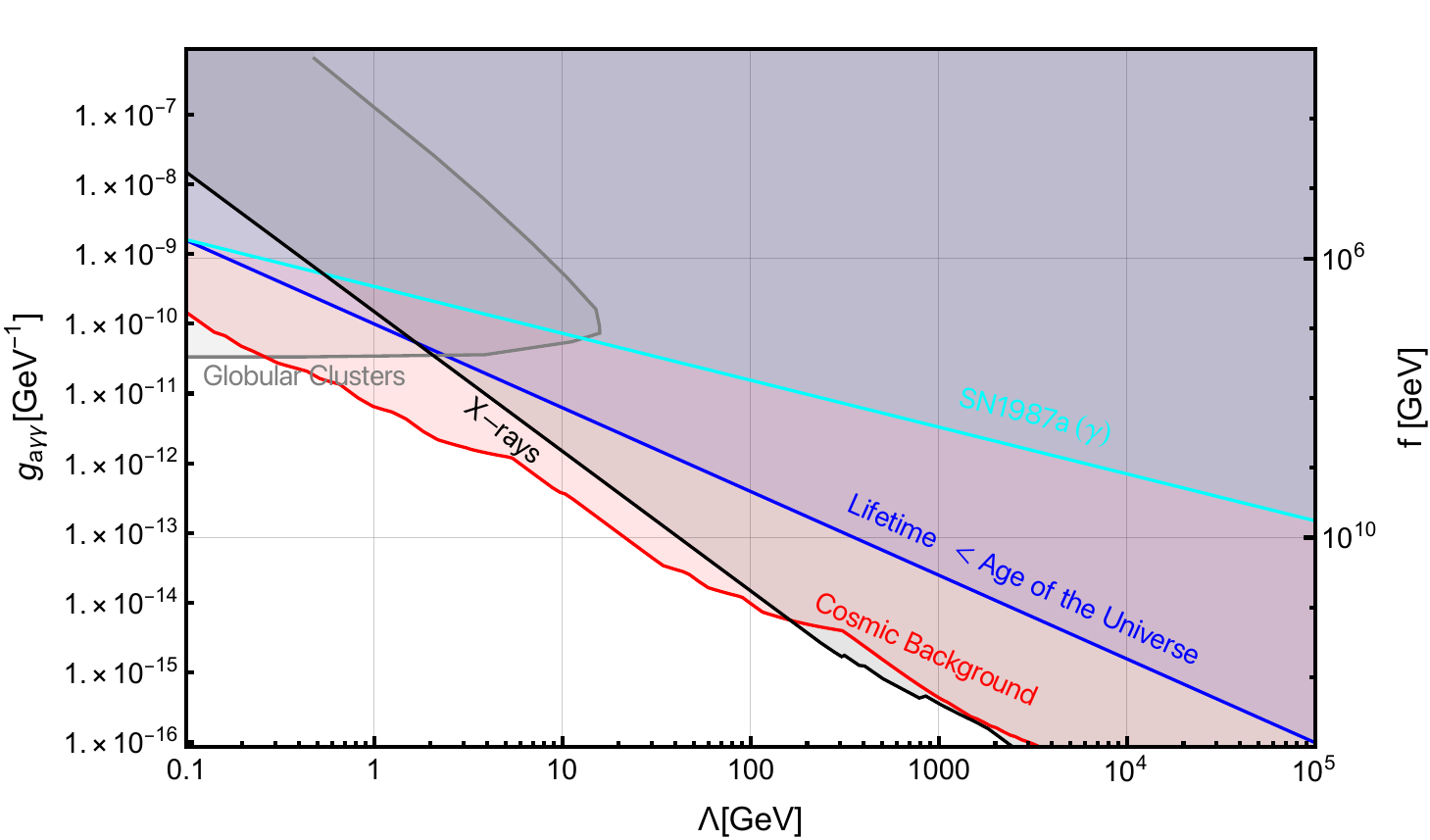}
    \caption{Existing astrophysical and cosmological constraints on the heavy 
	mass eigenstate assuming canonical coupling to photons. The bounds displayed are from 
	SN1987a~\cite{Jaeckel:2017tud,Hoof:2022xbe,Muller:2023vjm} (Cyan), 
	from the cosmic microwave
	background~\cite{Porras-Bedmar:2024uql} (red), from the $R$ and $R_2$ 
	parameters in Globular clusters~\cite{Ayala:2014pea,Dolan:2022kul} 
	(gray), and from X-ray observations~\cite{Cadamuro:2011fd} (black).} 
    \label{Fig: Lifetime_heavy}
\end{figure}

The axions (mass eigenstates) can decay into both standard model and other 
axions. Decays into other axions are suppressed in the case of large 
hierarchies between instanton scales~\cite{Gendler:2023kjt}, while decays into
standard model particles are highly model dependent. 
In many explicit construction of a particular string model, the coupling of 
the axions to QCD is natural~\cite{Demirtas:2021gsq}. The decay to gluons can 
be readily estimated and is negligible~\cite{Aloni:2018vki} for small masses,
but it can be significant for heavier axions. The couplings to fermions
are highly model dependent~\cite{Benabou:2023npn}, so we will only consider 
the decay to photons in the following. 
Depending on the masses of the eigenstates, the overlap between these axions 
and the QCD axion can be negligible~\cite{Gavela:2023tzu}, which can suppress 
the coupling to photons. However, for several explicit string theory 
constructions this coupling appears to be robust~\cite{Gendler:2023kjt}. The 
coupling of the flavor states to photons can be naturally parameterized by:
\begin{equation}
    C_\alpha\equiv 2\pi\frac{g_{a\gamma\gamma}f_a}{\alpha_{\text{EM}}},
\end{equation}
where $g_{a\gamma\gamma}$ is the axion coupling to photons and 
$\alpha_{\text{EM}}$ is the fine structure constant. It is often a good 
approximation to assume $C_\alpha\sim 1$.
The width associated to the decay of a mass eigenstate to two photons is then
given by:
\begin{equation}
\label{Eq: decay_rate_to_photon}
	\Gamma_{\phi_i \to \gamma \gamma} = \frac{M_{\phi_i}^3 
	g_{\phi_i \gamma \gamma}^2}{64 \pi}, \quad i=1,2,
\end{equation}
with the coupling constants:
\begin{equation}
\label{Eq: g_mass_eigenstates}
    g_{\phi_2 \gamma \gamma}^2=g_{a_1\gamma\gamma}^2\cos^2\theta+
	g_{a_2\gamma\gamma}^2\sin^2\theta,
	\quad  g_{\phi_1 \gamma \gamma}^2=g_{a_1\gamma\gamma}^2\sin^2\theta+
	g_{a_2\gamma\gamma}^2\cos^2\theta.
\end{equation}
\begin{figure}[t]
    \centering
    \includegraphics[width=0.85\textwidth]{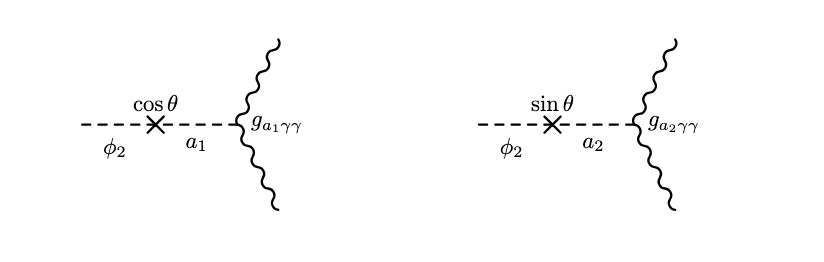}
    \caption{The decay of the heavier mass eigenstate $\phi_2$ to two photons. It can be clearly seen that the total decay width is proportional to
	the couplings given in eq.~\ref{Eq: g_mass_eigenstates}.}
    \label{Fig: Feynmann_diagram}
\end{figure}
Since the axion decay constants, $f_a$, are in general distributed rather 
narrowly~\cite{Gendler:2023kjt}, we assume that $f_1=f_2=f$ for concreteness. 
In this approximation, the coupling to photons is then given by:
\begin{equation}
\label{Eq: photon_coupling}
    g_{\phi_i\gamma\gamma}= \frac{\alpha_{\text{EM}}}{2\pi f}.
\end{equation}
In the limit $\Lambda\gg \Lambda_b$, we can write down the masses and the 
mixing angle:
\begin{equation}\label{Eq: limit_mass_angle}
\begin{split}
    M_{\phi_1}^2 &= \frac{ N_2^2 N_b^2}{\left(N_1^2+N_2^2\right)}\frac{\Lambda_b^4}{f^2}, \\
    M_{\phi_2}^2 &= (N_1^2+N_2^2)\frac{\Lambda^4}{f^2},\\
    \theta &=\frac{1}{2}\sin^{-1}{\left(\frac{2N_1N_2}{N_1^2+N_2^2}\right)}.
\end{split}
\end{equation}
With the canonical coupling to photons given by eq.~\ref{Eq: photon_coupling},
the lifetime of the heavier mass eigenstates turns out to be:
\begin{equation}
    \tau_{\phi_2}=\frac{256\pi^3 f^5}{(N_1^2+N_2^2)^{\frac{3}{2}}\alpha_{\text{EM}}^2\Lambda^6}=8.8\times 10^{20}\ \ \text{s}\left(\frac{f}{10^{10}\text{GeV}}\right)^5\left(\frac{10^2\text{GeV}}{\Lambda}\right)^6
\end{equation}
where we have used $N_1=2$ and $N_2=1$. These are the values that we use 
through the rest of this section. As can be seen from 
Fig.~\ref{Fig: Lifetime_heavy}, for some part of the parameter space the 
heavier mass eigenstate has a lifetime shorter than the age of the universe,
so they cannot constitute the dark matter in the universe. On the other
hand, the bulk of the currently unconstrained parameter space corresponds to 
the heavier axion having a lifetime that is much longer than the age of the 
universe, so it can potentially constitute the dark matter in the universe subject to other constraints from X-rays and CMB, as shown in Fig.~\ref{Fig: Lifetime_heavy}.
However, once we take into account the relic abundance, the viable part of
the parameter space is reversed. As we now show, the heavier eigenstate is
typically overproduced, which is excluded unless it decays fast enough.
If the heavy state is long lived compared to the annihilation time of the DWs, but short lived compared to the time of BBN, then it could dominate the energy density of the universe for a short period of time and then decay, leading to an EMD era.

Let us now, thus, discuss the axion abundances generated by different 
production mechanisms.
\subsubsection{Axions from the collapse of domain walls}

Moderately relativistic axions produced from the collapse of the network can 
serve as the cold dark matter. Let us first, for simplicity, discuss the 
production from the domain walls, ignoring for the moment the contribution 
of the strings.
In this case, the evolution of the energy density in domain walls 
$\rho_{\dw}$, axions $\rho_a$, and gravitational waves $\rho_{\gw}$ is 
given by the coupled differential equations: 
\begin{align}
	\dv{\rho_{\dw}}{t} &= -H\rho_{\dw} - 
	\eval{\dv{\rho_{\dw}}{t}}_{\text{emission}}, \\
	\dv{\rho_a}{t} &= -3 H\rho_{a}+ \dv{\rho_{\dw\rightarrow a}}{t}, \\
	\dv{\rho_\gw}{t} &= -4 H\rho_{\gw}+ \dv{\rho_{\dw\rightarrow \gw}}{t}, 
\end{align}
where 
\begin{equation}
	\eval{\dv{\rho_{\dw}}{t}}_{\text{emission}} = 
	\dv{\rho_{\dw\rightarrow a}}{t}+\dv{\rho_{\dw\rightarrow \gw}}{t}.
\end{equation}
From eqs.~\ref{Eq: energydensityDW} and~\ref{Eq: GWfromDW}, we find:
\begin{equation}
	\eval{\dv{\rho_{\dw}}{t}}_{\text{emission}}  = 
	\mathcal{A} \frac{\sigma_{\dw}}{2t^2},
\end{equation}
and:
\begin{equation}
	\dv{\rho_{\dw\rightarrow \gw}}{t} = 
	\frac{2 \epsilon_{\gw} G \mathcal{A}^2 
	\sigma_{\dw}^2}{t} .
\end{equation}
The total energy emmitted in the form of axions per unit comoving volume is 
given by:
\begin{equation}
	E_a(t) \equiv R^3(t)\rho_a(t)= \int_{t_i}^{t}{ \dd{t'} 
	R^3(t') \bqty{ \frac{\mathcal{A} \sigma_{\dw}}{2t'^2} 
	- \frac{2 \epsilon_{\gw} G \sigma_{\dw}^2 \mathcal{A}^2}{t'}} },
\end{equation}
where $t_i$ is the time when the wall begins to radiate axions, which
we can assume to be equal to the wall formation time. For $t\gg t_i$, the 
total energy in axions is then found to be: 
\begin{equation}
	E_a(t) \simeq R^3(t) \pqty{ \mathcal{A} \frac{\sigma_{\dw}}{t} - 
	\frac{4}{3} \epsilon_{\gw} G \sigma_{\dw}^2 \mathcal{A}^2}.
	\tag{4.12}
\end{equation}
The walls emit axions in the flavor state that they couple 
to~\cite{Benabou:2023npn}, which in our model is $a_1$. We can then
find the abundance of the different mass eigenstates at late times  
by projecting the mass eigenstates onto the flavor eigenstate $a_1$.
More explicitly:
\begin{equation}
    E_{\phi_2}(t)= E_a(t)\cos^2\theta \quad \text{and}\quad 
	E_{\phi_1}(t)= E_a(t)\sin^2\theta,
\end{equation}
where the mixing angle $\theta$ was obtained in eq.~\ref{Eq: rotation_angle}. 
We can relate the total
energy emitted in the form of axions to the width of the wall:
\begin{equation}
    \frac{E_{\phi_i}(t)}{N_{\phi_i}(t)}=\sqrt{1+\epsilon_a^2}
	\left(\frac{1}{\delta}\right),
\end{equation}
where $\delta = \frac{f_1}{N_1\Lambda^2}$ is the width of the wall, and 
$\epsilon_a$ can be obtained from simulations. The numerical experiments
in~\cite{Hiramatsu:2012sc,Kawasaki:2014sqa} show that $\epsilon_a$ has a mild 
dependence on the domain wall number and, while it increases slightly for 
large values of the domain wall number, it falls in the 
range $\epsilon_a\approx 1.2-1.5$. Once the domain walls have annihilated,
no more axions are being produced. If they are long lived, for times within
their lifetime, the number density of axions per comoving volume does not 
change:
\begin{equation}
    n_{\phi_i}(t_0) = 
	n_{\phi_i}(t_0)\left(\frac{R(t_{\text{ann}})}{R(t_0)}\right)^3.
\end{equation}
The axions emitted by the walls are mildly relativistic at the time of 
emission~\cite{Hiramatsu:2012sc,Kawasaki:2014sqa}, and they evolve to 
contribute to the population of cold axions today. The energy density in
the mass eigenstates today is:
\begin{equation}
    \rho_{\phi_i}(t_0)=n_{\phi_i}(t_0)M_{\phi_i},
\end{equation}
which corresponds to a fractional energy density:
\begin{equation}
\label{Eq: Omega_axion_DW}
	\begin{pmatrix}
		\Omega_{\phi_2,\text{DW}}(t_0) \\ 
		\Omega_{\phi_1,\text{DW}}(t_0)
	\end{pmatrix}
	=\frac{1}{\rho_c(t_0)}\frac{\delta}{\sqrt{1+\epsilon_a^2}}
	\pqty{ \mathcal{A} \frac{\sigma_{\dw}}{t_{\text{ann}}} 
	- \frac{4}{3} \epsilon_{\gw} G \sigma_{\dw}^2 \mathcal{A}^2}
	\left(\frac{R(t_{\text{ann}})}{R(t_0)}\right)^3 
	\times
	\begin{pmatrix}
		\cos^2\theta M_{\phi_2}\\
		\sin^2\theta M_{\phi_1}
	\end{pmatrix},
\end{equation}
where $\frac{R(t_{\text{ann}})}{R(t_0)}=\left(\frac{g_{*s0}}{g_{*,s,\text{ann}}}\right)^{1/3}\frac{T_0}{T_{\text{ann}}}$, 
and $H_0=1.4\times 10^{-33}\text{eV}$, $T_0=2.4\times 10^{-4}\text{eV}$ are 
the Hubble constant and the photon temperature today. 
Nominally, $\Omega_{\phi_i,\dw}$ in eq.~\ref{Eq: Omega_axion_DW} can become
negative if the network annihilation happens sufficiently late, 
$\Omega_{\dw}(t_{\text{ann}})>\frac{8\pi}{\epsilon_{\text{gw}}}>1$. 
In this case, eq.~\ref{Eq: Omega_axion_DW} breaks down, since the 
backreaction of the gravitational waves emitted by the domain walls becomes 
important~\cite{Hiramatsu:2012sc}. This regime is naturally avoided as long 
as the domain walls do not dominate the energy density of the universe. 
Indeed, when $\Omega_{\dw}(t_{\text{ann}})\ll 1$, it is a good approximation 
to ignore the negative contribution of the gravitational wave emission. 
In that case, using eq.~\ref{Eq: limit_mass_angle}, we find:
\begin{align}
\label{Eq: Omega2DW}
	&\Omega_{\phi_2,\dw}(t_0)\approx c_1
	\pqty{\frac{ f}{\mpl}}^{3/2}
	\frac{\Lambda ^3 T_0^3 \cos^2\theta}{\Lambda _b^2 \rho_{c,0}}
	=7.9\left(\frac{\Lambda}{10^2\text{GeV}}\right)^3
	\left(\frac{1\text{GeV}}{\Lambda_b}\right)^2
	\left(\frac{f}{10^{10}\text{GeV}}\right)^{\frac{3}{2}},\\
\label{Eq: Omega1DW}
	&\Omega_{\phi_1,\dw}(t_0)\approx c_2
	\left(\frac{ f}{\mpl}\right)^{3/2}
	\frac{\Lambda T_0^3\sin^2\theta}{ \rho_{c,0}}
	=3.9\times 10^{-5}
	\left(\frac{\Lambda}{10^2\text{GeV}}\right)
	\left(\frac{f}{10^{10}\text{GeV}}\right)^{\frac{3}{2}},
\end{align}
where 
$c_1=12\left(\frac{g_{*s,0}}{g_{*s}(T_{\text{ann}})}\right)\left(\frac{\left(N_1^2+N_2^2\right)\mathcal{A}^3 C_{\text{ann}}}{N_1 ^2g_*(T_{\text{ann}})^{3/2}(1+\epsilon _a^2)}\right)^{\frac{1}{2}}$ and $c_2 =c_1\left(\frac{N_2^2N_b^2}{(N_1^2+N_2^2)}\right)^{\frac{1}{2}}$ are dimensionless quantities. 
It should be noted that $c_1$ and $c_2$ have a mild dependence on 
$\Lambda$, $\Lambda_b$ and $f$, which we have neglected in the second part of 
eqs.~\ref{Eq: Omega2DW} and~\ref{Eq: Omega1DW}. 

\subsubsection{Axions from the misalignment mechanism}

In addition to the dark matter axions sourced by the domain wall network, 
the total axion dark matter abundance also includes a contribution from
the misalignment mechanism. Barring the presence of a large initial kinetic 
energy of the axion field, the zero mode remains frozen until the axion field 
evolution becomes underdamped, 
\begin{equation}
    M_{\phi_i}\simeq 3H(t_{\text{osc},i})
\end{equation}
where $t_{\text{osc},i}$ is the time when the $i$th mass eigenstate begins to
oscillate. Since the instanton potential is time independent, the masses
of the mass eigenstates $M_{\phi_i}$  are also time independent, unlike the 
case of the QCD axion. The total energy density in each mass eigenstate 
from misalignment is then given by:
\begin{equation}
    \rho_{\phi_i,\text{mis}}(t_0) = \frac{1}{2}M_{\phi_i}^2\langle
	\theta^2_{i,\text{mis}}\rangle f_{\phi_i}^2
	\left(\frac{R(t_{\text{osc}})}{R(t_0)}\right)^3
\end{equation}
where $\theta_i = \frac{a_i}{f_i}$. We are neglecting possible anharmonic 
effects in the axion field evolution that could be important if 
$\langle\theta_{i,\text{mis}}\rangle$ is close to $\pi$~\cite{Turner:1985si,Lyth:1991ub,Strobl:1994wk,Kobayashi:2013nva,Bae:2008ue,Visinelli:2009zm}. 
The initial misalignments of the mass eigenstates are:
\begin{equation}
\langle\theta^2_{2,\text{mis}}\rangle f_{\phi_2}^2 \approx 
\frac{\pi^2}{3}(f_1\cos\theta+f_2\sin\theta)^2,  \quad 
	\langle\theta^2_{1,\text{mis}}\rangle f_{\phi_1}^2 \approx
	\frac{\pi^2}{3}(f_1\sin\theta+f_2\cos\theta)^2,
\end{equation}
where we have approximated the initial values $a_i/f_i$ to be uniformly distributed. We further elaborate on this aspect in 
App.~\ref{App: Initial_Misalignment}.
Taking the benchmark values $f_1=f_2=f$, we find the fractional axion
energy density from misalignment:
\begin{align}
\label{Eq: Misalignment_Omega2}
&\Omega_{\phi_2,\text{mis}}(t_0)\approx c_3\left(\frac{ f}{\mpl}\right)^{3/2}\frac{T_0^3\Lambda}{\rho_{c,0}}=0.42\left(\frac{\Lambda}{10^2\text{GeV}}\right)\left(\frac{f}{10^{10}\text{GeV}}\right)^{\frac{3}{2}}\\
\label{Eq: Misalignment_Omega1}
&\Omega_{\phi_1,\text{mis}}(t_0)\approx c_4\left(\frac{ f}{\mpl}\right)^{3/2}\frac{T_0^3\Lambda_b}{\rho_{c,0}}=1.9\times 10^{-4}\left(\frac{\Lambda_b}{1\text{GeV}}\right)\left(\frac{f}{10^{10}\text{GeV}}\right)^{\frac{3}{2}},
\end{align}
where $c_3=1.6\left(\frac{g_{*s,0}}{g_{*s,2}}\right)\frac{(N_1+N_2)^2g_{*,2}^{3/4}}{(N_1^2+N_2^2)^{3/4}}$ and $c_4=1.6\left(\frac{g_{*s,0}}{g_{*s,1}}\right)\frac{(N_1+N_2)^2g_{*,1}^{3/4}}{(N_1^2+N_2^2)^{3/4}}(N_2^2N_b^2)^{1/4}$ are 
dimensionless coefficients, and $g_{*,i}=g_*|_{M_{\phi_i}=3H}$ is evaluated
at the time when the mass eigenstates start to oscillate. Once more, we have
neglected the mild dependence of $c_3$ and $c_4$ on $\Lambda,\Lambda_b,f$
that enters through $g_{*,i}$.

\subsubsection{Axion emission from strings in the scaling regime}
\label{subsection: scaling_axions}

Most of the numerical studies of axion emission from field theory strings 
are performed in the context of $N_\dw=1$ models with unstable domain walls~\cite{Harari:1987ht,Hagmann:1990mj,Davis:1989nj,Battye:1993jv,Battye:1994au,Vilenkin:1982ks,Sikivie:1982qv,Davis:1986xc,Shellard:1987bv,Klaer:2017ond,Vaquero:2018tib,Gorghetto:2018myk,Gorghetto:2020qws,Buschmann:2019icd,Buschmann:2021sdq}. 
In this case, the results of the simulations are consistent with the 
assumption that the network of strings emits relativistic axions with a 
spectrum $\dd{E_a}/\dd{\omega}\sim 1/\omega$.
On the other hand, in the case of a stable string domain wall network in 
models with $N_\dw>1$, existing numerical simulations generally introduce
a bias potential that causes the network to decay at a time when the energy 
of the strings is subdominant compared to that of the domain 
walls~\cite{Hiramatsu:2012sc}. 
As a result, the bulk of the axion emission is constituted by the mildly
relativistic axions produced by the domain walls, while the strings in
the scaling regime merely contribute a residual long tail to the spectrum.


The axion emission from strings in a string theory axiverse scenario with flat
extra dimensions was studied in~\cite{Benabou:2023npn}. In this case, the 
rate of energy emission in mass eigenstates is related to the energy emitted
in flavor states by:
\begin{equation}
	\dv{E_{\phi_1}}{t}=\dv{E_{a_1}}{t}\sin^2\theta
	\quad
	\dv{E_{\phi_2}}{t}=\dv{E_{a_1}}{t}
	\cos^2\theta
\end{equation}
where $E_{a_1}$ and $E_{\phi_i}$ are the energy emitted in the flavor state 
$a_1$ and in the mass eigenstate $\phi_i$ respectively; 
and $\theta$ is the mixing angle defined in eq.~\ref{Eq: rotation_angle}. 
The total power emitted per unit volume is found by adding the contributions 
of loops of different sizes present in the network: 
\begin{align}
	\Gamma_{\phi_i}(t)=\int\dv{E_{\phi_i}}{t}n(\ell, t) \dd{\ell}.
\end{align}
Assuming an axion spectrum of the form  $\dv{\dot{E}_a}{\omega}\sim 1/\omega$,
we obtain the number density of axions emitted until time $t$:
\begin{align}
\label{Eq: n_phi_2}
    n_{\phi_2}(t)\approx \frac{4}{3}\sqrt{\frac{2}{\pi}}\frac{r_a\alpha}{\sqrt{r \kappa}}Hf\frac{\sqrt{f\mpl}}{\log\left(\frac{\mpl}{\Lambda}\right)}\cos^2\theta.
\end{align}

The characteristic energy of the axions emitted at time $t$ is given by 
$\omega(t)\sim H\mpl/f$. Then, we can assume that a particular
mass eigenstate effectively stops being emitted when 
$\omega(t_{\phi_i})\approx M_{\phi_i}$ (where we have implicitly defined the 
time $t_{\phi_i}$ by this equation), even though the string domain wall 
network 
persists long afterwards. This is particularly relevant for the heavier mass 
eigenstate $\phi_2$, as we will see that, for a large part of parameter space, 
the network disappears long after the strings are unable to produce $\phi_2$, 
$\tann\gg t_{\phi_2}\approx \mpl/\Lambda^2$. This is in contrast with the 
lighter mass 
eigenstate $\phi_1$, for which $t_{\phi_1}\geq \tann$ as long as the network 
annihilates before it dominates the energy density of the universe. 

Putting this together, we obtain the total abundance of $\phi_2$ from strings
at present: 
\begin{align}
	\Omega_{\phi_2, \text{str}}(t_0) &=
	\frac{n_{\phi_2}(t_{\phi_2})M_{\phi_2}}{\rho_{c,0}}
	\left(\frac{R(t_{\phi_2})}{R(t_0)}\right)^3
	\nonumber \\
	&\approx c_5\left(\frac{ f}{\mpl}\right)^{\frac{1}{2}}
	\frac{\Lambda T_0^3 \cos^2\theta}{\rho_{c,0}}
	\nonumber \\
	&=0.97\times 10^2\left(\frac{\Lambda}{10^2\text{GeV}}\right)
	\left(\frac{f}{10^{10}\text{GeV}}\right)^{\frac{1}{2}},
	\label{Eq: Omega2_string}
\end{align}
where $c_5=0.27\ \frac{4}{3}\sqrt{\frac{2}{\pi}}\frac{r_a\alpha}{\sqrt{r \kappa}}\left(\frac{g_{*s0}}{g_{*s}(t_{\phi_2})}\right)\frac{(N_1^2+N_2^2)^{1/4}}{g_{*}(t_{\phi_2})^{3/4}}$. 

\begin{figure}[t]
    \centering
    \includegraphics[width=0.85\textwidth]{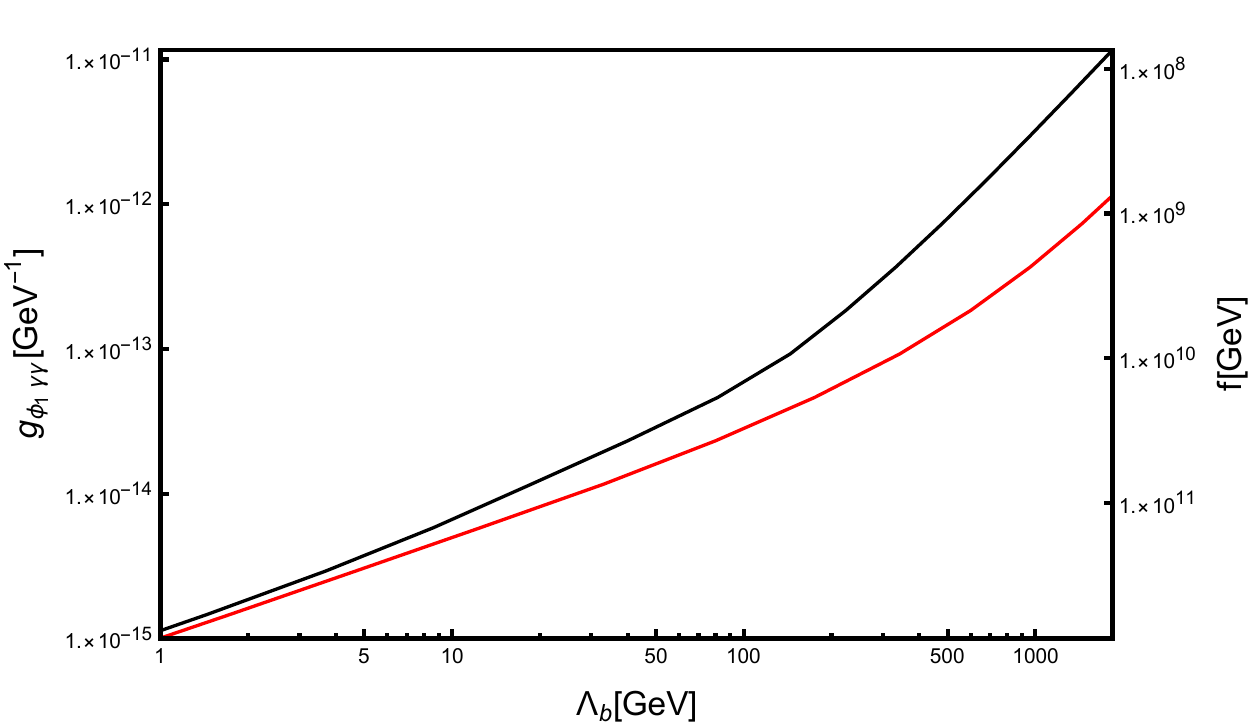}
    \caption{The canonical ALP to photon coupling for which the light mass
	eigenstat $\phi_1$ can make the entirety of the dark matter. The black
	(red) curve corresponds to the case of domain walls with an
	energy density of $10^{-2}\rho_{\text{tot}}$ 
	$(10^{-4}\rho_{\text{tot}})$ at the time of the annihilation of the 
	network.}
    \label{Fig: phi_1_DM_saturation}
\end{figure}

Now we can calculate the total abundance of the heavy mass eigenstate, 
$\phi_2$ by adding the contribution from the domain walls, the strings and 
from the misalignment mechanism. We see at once from eqs.~\ref{Eq: Omega2DW},
~\ref{Eq: Misalignment_Omega2} and~\ref{Eq: Omega2_string}, that the heavier 
mass eigenstate is typically overproduced unless both the instanton scale and 
the axion decay constant are small. 
If the string-domain wall network annihilates before dominating the energy 
density of the universe, the lighter mass eigenstate can be produced by the 
strings up until the time annihilation of the network. 
The abundance of $\phi_1$ is given by:
\begin{align}
	\Omega_{\phi_1, \text{str}}(t_0)&=
	\frac{n_{\phi_1}(\tann)M_{\phi_1}}{\rho_{c,0}}
	\left(\frac{R(\tann)}{R(t_0)}\right)^3 
	\nonumber\\
	&\approx c_6\left(\frac{ f}{\mpl}\right)\frac{\Lambda T_0^3 
	\sin^2\theta}{\rho_{c,0}\log\left(\frac{\mpl}{\Lambda_b}\right)}
	\nonumber\\
	&=3.2\times 10^{-4}\left(\frac{\Lambda}{10^2\text{GeV}}\right)
	\left(\frac{f}{10^{10}\text{GeV}}\right),
	\label{Eq: Omega1_string}
\end{align}
where $c_6=0.76 \ \frac{4}{3}\sqrt{\frac{2}{\pi}}\frac{r_a\alpha}{\sqrt{r \kappa}}\left(\frac{g_{*s0}}{g_{*s}(\tann)}\right)\left(\frac{N_2^2N_b^2\mathcal{A} C_{\text{ann}}}{\left(N_1^2+N_2^2\right)g_*(\tann)^{3/2}}\right)^{\frac{1}{2}}$. 
Adding the contributions from misalignment~\ref{Eq: Misalignment_Omega1}, 
the domain walls~\ref{Eq: Omega1DW}, and from the 
strings~\ref{Eq: Omega1_string}, we conclude that the light mass eigenstate, 
$\phi_1$ can constitute the dark matter in the universe. 
We can see this fact explictly in fig.~\ref{Fig: phi_1_DM_saturation},
which shows the values of $\Lambda_b$ vs
$f$ for which $\phi_1$ saturates the observed dark matter abundance.
We should keep in mind that we have assumed a standard cosmological
evolution throughout, but, as we pointed out earlier,
the presence of the heavier mass eigenstate, $\phi_2$, can trigger an epoch of
early matter domination. However, since an EMD epoch would dilute the 
abundance of cold $\phi_1$s relative to a radiation dominated expansion, 
we see matter domination can not lead to an overabundance of $\phi_1$. 
We further elaborate on this EMD scenario in App.~\ref{App: EMD}, where 
we show that $\phi_1$ can saturate the cosmological dark matter abundance also
in that case. We, thus, conclude that the abundance of the light axion mass 
eigenstate resulting from the decay of the network and from the misalignment
mechanism can account for the dark matter in the universe.
 
\section{Conclusions}
\label{sec: conclusion}
In this article, we have explored the phenomenology of
axionic string-domain wall networks in the context of string theory.
In scenarios containing several axions, as in axiverse constructions,
multiple instantons generate a bias term in the potential, which leads
to the decay of the network and avoids the domain wall problem.

The decay of the network generates interesting observable signatures,
which have been extensively studied in the context of field theory axions.
But, an important feature of string theory axion constructions with flat 
extra dimensions is that the tension of the strings in the network
is parametrically larger than for field theory axions. Importantly, this
fact can lead to a distinct phenomenology that can enable testing these
scenarios in current and forthcoming experiments.

For instance, the spectrum of gravitational waves produced by the decay 
of the network has a characteristic ``double peak'' structure, that arises
because the contribution of the strings in the network, which due to their
higher tension cannot be neglected, dominates at higher frequencies. The
combined shape of the spectrum is different from the predictions of other
early universe phenomena such as phase transitions, field theory axions.
Importantly, this singular spectrum provides a telltale sign of string
theory axions in gravitational wave observations.

Furthermore, working with a simplified model with two axions, we have shown 
that the heavier one is typically overproduced. For adequate values of the 
two-photon coupling, it decays fast enough to avoid conflict with BBN or 
overclosing the universe. Importantly, the lighter eigenstate can then account
for the dark matter in the universe. 

Our work is motivated by string theory constructions of axions with flat
extra dimensions, and although some of our conclusions might not extend to
the case of warped extra dimensions, some might. We have also left aside
the study of other interesting phenomenology that could take place in the
early universe, such as the production of primordial black holes. These
are interesting directions for future work.



\section*{Acknowledgments}

We thank J.J. Blanco-Pillado, D. Dunksy, J. Garriga, O. Pujol\`as,
F. Rompineve and A. Vikman for useful discussions. FF would like to thank the 
University of the Basque Country (UPV/EHU) for its hospitality during the 
Cosmological Olentzero Workshop.

\appendix
\section{Condition on the charges}
\label{App: charge_condition}
In the main text, we have assumed that the bias potential only couples to the flavor state $a_1$:
\begin{equation}
    V_b(a_1) =\frac{\Lambda_b^4}{2}
	\left[1- \cos\left(N_b\frac{a_1}{f_1}\right) \right].
\end{equation}
However, in general the bias potential can couple to both $a_1$ and $a_2$. 
Assuming that the two potentials are of the form:
\begin{equation}
    V_1(a_1,a_2) = \Lambda^4\left[1-\cos\left(N_{11}\frac{a_1}{f_1}+N_{12}\frac{a_2}{f_2}\right)\right],
\end{equation}
\begin{equation}
    V_2(a_1,a_2) = \Lambda_b^4\left[1-\cos\left(N_{21}\frac{a_1}{f_1}+N_{22}\frac{a_2}{f_2}\right)\right],
\end{equation}
we can derive the condition for which $V_2$ allows for a unique 
minimum~\cite{Benabou:2023npn}.
If the charge matrix is invertible, the number of minima of the potential in $a_1$ is given by"
\begin{equation}
    N_{\text{DW}}=\frac{|N_{11}N_{22}-N_{12}N_{21}|}{|\text{gcd}(N_{22},N_{12})|}.
\end{equation}
We see that, for our choice of values, namely, $N_{11}=2$, $N_{12}=N_{21}=1$ and $N_{22}=0$, $\NDW=1$ where $\text{gcd}(0,b) = b$ for positive integer $b$.
\section{Role of unequal initial abundance of vacua}
\label{App : Unequal_abundance}
Since the different minima are non completely degenerate, the axion field does
not roll down to two nearby minima with equal probability. The true lower 
energy vacuum is more likely to be populated. This asymmetry in the number of 
different domain walls acts as a different bias in the network, which can 
also lead to the annihilation of the string domain wall network. 
In the case of $N_{\text{DW}}=2$ with two non-degenerate minima, the field 
value has a higher probability of being in the true minimum than in the false 
minimum at the time of domain wall formation. The ratio of the two 
probabilities is given by~\cite{Hiramatsu:2010yn}:
\begin{equation}
    \frac{p_f}{p_t}=\exp\left(-\frac{\Delta V_b}{V_{\text{min}}}\right),
\end{equation}
where $p_f$ and $p_t$ are the probabilities for the field value to fall into the false and true minimum respectively, $\Delta V_b$ in the difference in potential between the two minima, and $V_{\text{min}}$
is the minimum value of the leading potential. If we measure the extent of 
the unequal occupancy by $p_f = 0.5-\epsilon$, the numerical simulations 
in~\cite{Larsson:1996sp} found:
\begin{equation}
\label{Eq:bias_annihilation}
    \frac{\tau_{\text{ann}}}{\tau_{\text{form}}} \simeq 
	\epsilon^{-\frac{D}{2}},
\end{equation}
where $\tau_{\text{ann}}$ denotes the conformal time of domain wall 
annihilation, $\tau_{\text{form}}$ is the conformal time of domain wall 
formation, and $D$ is the spatial dimension. Even though, the analysis 
of~\cite{Coulson:1995nv, Larsson:1996sp} considered the specific case 
of $N_{\text{DW}}=2$, we assume the rate of annihilation is not 
parametrically faster for the case of  $N_{\text{DW}}>2$ than 
eq.~\ref{Eq:bias_annihilation}. In this case, we have: 
\begin{equation}
    \frac{\tau_{\text{ann}}}{\tau_{\text{form}}} \simeq \left(\frac{2 V N_{\text{DW}}}{\Delta V_b}\right)^3 t_{\text{form}}{\frac{3}{2}},
\end{equation}
where we have used $D=3$ spatial dimensions. Now we can estimate the 
typical annihilation timescale associated to the unequal vacuum abundance. 
During radiation domination, $\tau\sim\sqrt{t}$. Parametrically, 
\begin{equation}
    t_{\text{ann, unequal abundance}} = \left(\frac{2 V N_{\text{DW}}}{\Delta V_b}\right)^3 t_{\text{form}} \sim \frac{8N_{\text{DW}}^2f\Lambda^{10}}{\Lambda_b^{12}}.
\end{equation} 
This timescale is much larger than the one associated to the pressure force,
which is given in~\ref{Eq: tann}. Hence, our assumption to just consider
the annihilation due to the pressure force is justified.

\section{GWs from decays during an early matter dominated era}
\label{App:emd}
If the collapse happens during matter domination, the peak amplitude of the GW estimate changes \cite{Saikawa:2017hiv}
\begin{equation}
	\Omega_{\gw}h^2(t_0) = \Omega_{\text{rad}}h^2 
	\left(\frac{g_*(T_{\text{reh}})}{g_{*0}}\right) \left(\frac{g_{*s0}}{g_{*s}(T_{\text{reh}})}\right)^{4/3} \left(\frac{H_{\text{reh}}}{H_{\text{ann}}}\right)^{2/3} \Omega_{\gw}(t_{\text{ann}}),
\end{equation}
where $T_{\text{reh}}$ is the reheating temperature, and $H_{\text{reh}}$ and $H_{\text{ann}}$ are the Hubble parameters at $T = T_{\text{reh}}$ and $T = T_{\text{ann}}$, respectively. In the case where the reheating is under perturbative control~\cite{Giudice:2000ex}, the Hubble parameter at this stage is of the matter dominated era is given by:
\begin{equation}
    H(T) = \left[\frac{5\pi^2 g_*^2(T)}{72 g_*(T_{\text{reh}})}\right]^{1/2} \frac{T^4}{M_P T_{\text{reh}}^2}.
\end{equation}
The peak frequency and the cutoff frequency also change in the following way:
\begin{equation}
    f_{peak} = H(t_{\text{ann}})\frac{R(t_{\text{ann}})}{R(t_0)}= H(t_{\text{ann}})\left(\frac{H_{\text{reh}}}{H_{\text{ann}}}\right)^{2/3}\frac{T_{\text{reh}}}{T_0}
\end{equation}

\begin{equation}
    f_{\delta} = \delta^{-1}\frac{R(t_{\text{ann}})}{R(t_0)}= \delta^{-1}\left(\frac{H_{\text{reh}}}{H_{\text{ann}}}\right)^{2/3}\frac{T(t_{\text{ann}})}{T_0}
\end{equation}

The resulting spectrum is similar to the standard case shown 
in~\fig{ perturbation}, but shifted to lower frequencies.

\section{Averaged misalignment angle}
\label{App: Initial_Misalignment}
Assuming that the angular variables $a_i/f_i$ are uniformly distributed, we get the following for the initial misalignment value of the mass eigenstate:
\begin{equation}
\langle\theta_{2,\text{mis}}\rangle^2 f_{\phi_2}^2 \approx 
\frac{\pi^2}{3}(f_1\cos\theta+f_2\sin\theta)^2,  \quad 
	\langle\theta_{1,\text{mis}}\rangle^2 f_{\phi_1}^2 \approx
	\frac{\pi^2}{3}(f_1\sin\theta+f_2\cos\theta)^2,
\end{equation}
However, there are two important subtleties which we glossed over in the main 
text. Namely, if the initial misalignments of the flavor states are given 
by $a_i/f_i=\theta_i$ in a patch of the universe, the displacement of the mass eigenstate is:
\begin{align}
    \phi_2 = a_1 \cos\theta + a_2 \sin\theta =f_1\theta_1\cos\theta + f_2 \theta_2\sin\theta
\end{align}
Then, the ensemble average assuming a uniform distribution of $\theta_i$ 
yields:
\begin{align}
	\langle \phi_2^2 \rangle &
	\equiv \langle\theta^2_{2,\text{mis}}\rangle f_{\phi_2}^2 
	\nonumber\\
	&= \frac{\pi^2}{3}(f_1^2\cos^2\theta + f_2^2\sin^2\theta+2\cdot 
	\frac{3}{4}f_1f_2\sin\theta\cos\theta)
	\nonumber\\
	&\neq \frac{\pi^2}{3}(f_1\cos\theta+f_2\sin\theta)^2.
\end{align}
This has a negligible effect on the total abundance of dark matter when 
combined with the contribution from the strings and the domain walls.
However, a careful analysis of the misalignment mechanism needs to account for the non-uniform initial distribution caused by the time independent axiverse potential, which leads 
to the fields being more likely to populate the minima of the potential 
than other values in the field space. We leave this for a future analysis.

\section{No EMD from string network decay during scaling}
\label{App: No_EMD}
In this section we show that the heavier axion mass eigenstates produced from 
the strings can only dominate the energy density of the universe close to the 
annihilation time, $\tann$. The total energy density of the axions produced 
from the strings in the scaling regime is~\cite{Benabou:2023npn}:
\begin{align}
    \rho_{\phi_2}\approx \frac{16 r_a (8\pi)^{3/2} \alpha}{3 r^{3/2} \kappa^{3/2}} H^2 \sqrt{f_a M_{\text{pl}}^3}\log\left(\frac{\mpl}{\Lambda}\right)\cos^2\theta.
\end{align}
Assuming that the axions are emitted with a spectrum 
$\dd{\dot{E_a}}/\dd{\omega}\sim 1/\omega$, the number density of $\phi_2$ is given by eq.~\ref{Eq: n_phi_2}. Thus, we see that the mean energy of the axions at time $t$ is given by:
\begin{align}
\label{Eq: t_phi_2}
    \omega_{\phi_i}(t)\approx H(t)\frac{\mpl}{f}\log^2\left(\frac{\mpl}{\Lambda}\right).
\end{align}
The heavy mass eigenstate production stops before the annihilation of the network~\cite{Benabou:2023npn}, at time 
$\omega_{\phi_i}(t_{\phi_2})\approx M_{\phi_2}$. The $\phi_2$'s are 
non-relativistic at $t=t_{\phi_2}$. Hence, the total energy density in
$\phi_2$ at that time is given by:
\begin{align}
    \rho_{\phi_2}(t_{\phi_2})=n_{\phi_2}(t_{\phi_2})M_{\phi_2}\approx (N_1^2+N_2^2)\frac{\Lambda^4}{\log^3\left(\frac{\mpl}{\Lambda}\right)}\sqrt{\frac{f}{\mpl}}.
\end{align}
The fractional energy density in $\phi_2$ at that time is given by:
\begin{align}
    \Omega_{\phi_2}(t_{\phi_2})=\frac{ \rho_{\phi_2}(t_{\phi_2})}{3H(t_{\phi_2})^2\mpl^2}\approx\sqrt{\frac{f}{\mpl}}\frac{\log\left(\frac{\mpl}{\Lambda}\right)}{3}.
\end{align}
We, thus, see that at $t_{\phi_2}$ the universe is not $\phi_2$ dominated. 
If the universe is radiation dominated at $t_{\phi_2}$, we can evaluate the 
fractional energy density of $\phi_2$ at $t=\tann$ by assuming the universe 
remains radiation dominated until $\tann$. We will see that this assumption 
only breaks down close to $t=\tann$, thus serving as a reasonable 
approximation. The fractional energy density of $\phi_2$ at $t=\tann$:
\begin{align}
    \Omega_{\phi_2}(\tann)\sim\Omega_{\phi_2}(t_{\phi_2})\sqrt{\frac{\tann}{t_{\phi_2}}}\sim \frac{f}{\mpl}\frac{\Lambda^2}{\Lambda_b^2}\log\left(\frac{\mpl}{\Lambda}\right),
\end{align}
where we have used eq.~\ref{Eq: tann} for $\tann$ and eq.~\ref{Eq: t_phi_2} for $t_{\phi_2}$.
Applying eq.~\ref{Eq: bias_minimum}, so that the domain walls do not dominate the energy density of the universe, we find:
\begin{align}
    \Omega_{\phi_2}(\tann)<\log\left(\frac{\mpl}{\Lambda}\right).
\end{align}

\section{Dark matter abundance from EMD}
\label{App: EMD}
\begin{figure}[t]
    \centering
    \includegraphics[width=0.85\textwidth]{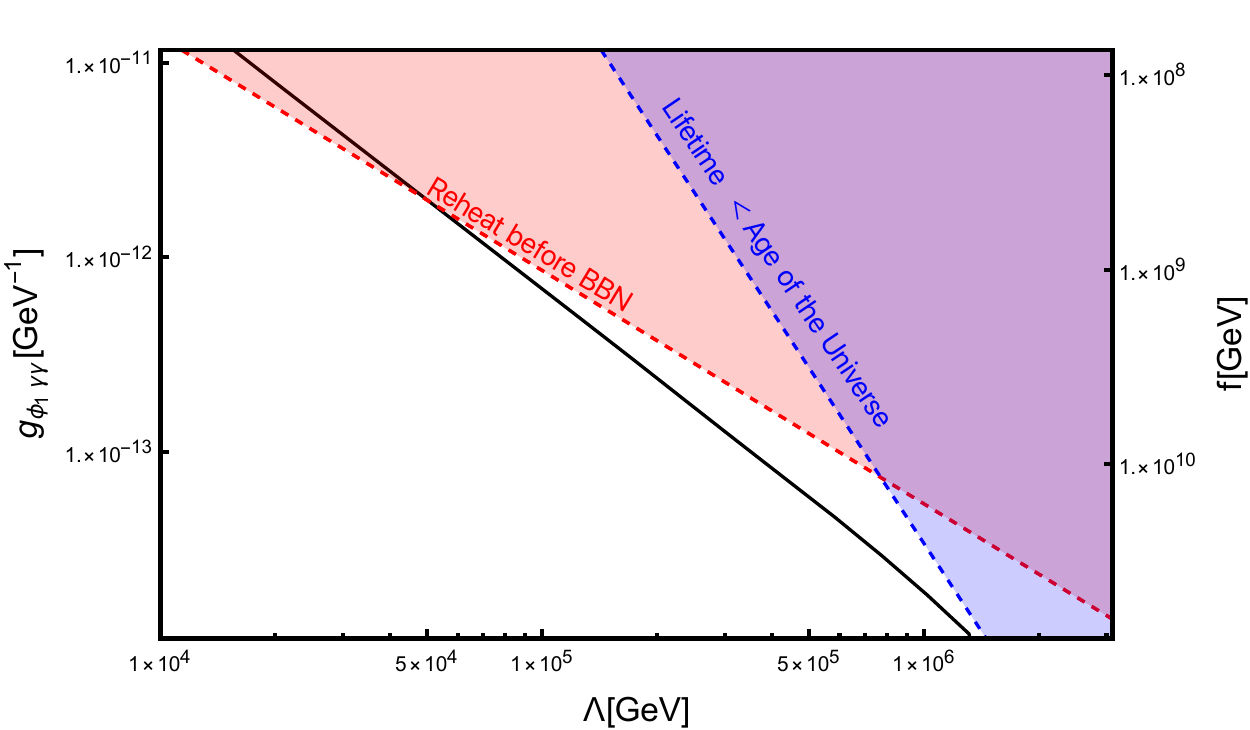}
    \caption{The canonical ALP to photon coupling for which the light ALP
	constitutes the entirety of the dark matter including a period of 
	early matter domination due to $\phi_2$. The black curve corresponds 
	to the case of domain walls being $10^{-1}\rho_{\text{tot}}$ at the 
	time of annihilation of the network. The red dashed line corresponds 
	to a reheating temperature of 5 MeV, allowing conventional BBN. 
	Assuming canonical coupling to photons, are bounds not shown are~\cite{Jaeckel:2017tud,Hoof:2022xbe,Muller:2023vjm,Porras-Bedmar:2024uql,Ayala:2014pea,Dolan:2022kul,Cadamuro:2011fd}.}
    \label{Fig: phi_1_DM_saturation_EMD}
\end{figure}

As shown in section~\ref{subsection: axionic_DM}, for the majority of 
parameter space the heavier eigenstate $\phi_2$ is overproduced
compared to the observed abundance of dark matter. If $\phi_2$ does not decay 
immediately, it can lead to a matter dominated era. The total decay width of 
$\phi_2$ has large uncertainties due to the model dependence of the couplings 
to SM fermions. The couplings to other axions are relatively 
suppressed~\cite{Gendler:2023kjt}. As a benchmark, we assume that the decay to photons dominates the total decay. The reheating temperature in that case 
satisfies:
\[
\rho_{\text{tot}}(T_{\text{RH}}) = \frac{\pi^2}{30} g_*(T_{\text{RH}}) T_{\text{RH}}^4 = \frac{4}{3} M_{\text{pl}}^2 \Gamma_{\phi_2\gamma\gamma}^2,
\]
where $\Gamma_{\phi_2\gamma\gamma}$ is given in eq.~\ref{Eq: decay_rate_to_photon}. In section~\ref{subsection: axionic_DM} we noted that, ignoring the EMD, 
the total abundance of $\phi_1$ can saturate the observed cosmological dark 
matter. We now discuss the impact of an EMD on that scenario. The mass of the 
axion eigenstates is temperature independent. The total energy density in 
$\phi_1$ today is given by:

\begin{align}
    \rho_{\phi_1,{\dw}}(t_0)=\rho_{\phi_1,{\dw}}(\tann)\left(\frac{R(\tann)}{R(\tEMD)}\right)^3\left(\frac{R(\tEMD)}{R(\tRH)}\right)^3\left(\frac{R(\tRH)}{R(t_0)}\right)^3,
\end{align}
where we have simplified the cosmological evolution by assuming the universe transitions directly from radiation domination to matter domination. Here, 
$\tEMD$ denotes the onset of matter domination which can be estimated as 
follows. The production of $\phi_2$ from the strings stops before the 
collapse of the nework, as shown in App.~\ref{App: No_EMD}. Assuming that
the universe is radiation dominated at this time, the fractional energy 
density of $\phi_2$ scales as:
\begin{align}
    \Omega_{\phi_2}(\tEMD)\sim\Omega_{\phi_2}(t_{\phi_2})\sqrt{\frac{\tEMD}{t_{\phi_2}}}\sim 1.
\end{align}
The Hubble at the onset of matter domination is:
\begin{align}
    H(\tEMD)\approx \frac{2048 \pi  \alpha r_a^2}{27 \kappa^2 r^2}\frac{f^2}{\mpl^2}M_{\phi_2}.
\end{align}
The scale factor during the matter dominated era behaves as:
\begin{align}
    \left(\frac{R(\tEMD)}{R(\tRH)}\right)=\left(\frac{H(\tEMD)}{H(\tRH)}\right)^{-\frac{2}{3}}.
\end{align}
Now we can evaluate the dark matter contribution from the misalignment 
mechanism. The $\phi_1$ field starts coherent oscillations when $ M_{\phi_1}\simeq 3H(t_{\text{osc},1})$. The oscillations start before matter domination:
\begin{align}
    \frac{H(\tEMD)}{H(t_{\text{osc},1})}\sim \frac{f^2\Lambda^2}{\mpl^2\Lambda_b^2}<\frac{f}{\mpl},
\end{align}
where in the last part, we have assumed that the bias potential is large enough so that the domain walls annihilate before they dominate the energy density of the universe, $\Lambda_b >\Lambda \sqrt{f/\mpl}$. The total abundance of $\phi_1$ from the misalignment mechanism is given by:
\begin{align}
    \rho_{\phi_1,{\text{mis}}}(t_0)=\rho_{\phi_1,{\text{mis}}}(\tosc)\left(\frac{R(\tosc)}{R(\tEMD)}\right)^3\left(\frac{R(\tEMD)}{R(\tRH)}\right)^3\left(\frac{R(\tRH)}{R(t_0)}\right)^3
\end{align}
The contribution from the strings can be calculated similarly by tracking the number density of $\phi_1$'s. Assuming that the universe is radiation dominated during the entirety of the network evolution,
\begin{align}
    \rho_{\phi_1,{\text{str}}}(t_0)=\rho_{\phi_1,{\text{str}}}(\tann)\left(\frac{R(\tann)}{R(\tEMD)}\right)^3\left(\frac{R(\tEMD)}{R(\tRH)}\right)^3\left(\frac{R(\tRH)}{R(t_0)}\right)^3.
\end{align}
Hence, the fractional energy density in $\phi_1$ today is given by
\begin{align}
    \Omega_{\phi_1, \text{tot}}(t_0)=\frac{\rho_{\phi_1,{\dw}}(t_0)+\rho_{\phi_1,{\text{mis}}}(t_0)+\rho_{\phi_1,{\text{str}}}(t_0)}{\rho_{c,0}}
\end{align}
where $\rho_{c,0}$ is the critical energy density today. In Fig.~\ref{Fig: phi_1_DM_saturation_EMD}, we see that $\phi_1$ can be the entirety of the 
cosmological dark matter. It should be noted that, although we have assumed 
a canonical coupling to photons, all of the production mechanisms for the 
axions are independent of such coupling. 
\bibliographystyle{JHEP}
\bibliography{Domain_walls.bib}

\providecommand{\href}[2]{#2}\begingroup\raggedright\begin{thebibliography}{100}

\bibitem{Witten:1984dg}
E.~Witten, {\it {Some Properties of O(32) Superstrings}},  {\em Phys. Lett. B} {\bf 149} (1984) 351--356.

\bibitem{Choi:1985je}
K.~Choi and J.~E. Kim, {\it {Harmful Axions in Superstring Models}},  {\em Phys. Lett. B} {\bf 154} (1985) 393. [Erratum: Phys.Lett.B 156, 452 (1985)].

\bibitem{Barr:1985hk}
S.~M. Barr, {\it {Harmless Axions in Superstring Theories}},  {\em Phys. Lett. B} {\bf 158} (1985) 397--400.

\bibitem{Svrcek:2006yi}
P.~Svrcek and E.~Witten, {\it {Axions In String Theory}},  {\em JHEP} {\bf 06} (2006) 051, [\href{http://arxiv.org/abs/hep-th/0605206}{{\tt hep-th/0605206}}].

\bibitem{Arvanitaki:2009fg}
A.~Arvanitaki, S.~Dimopoulos, S.~Dubovsky, N.~Kaloper, and J.~March-Russell, {\it {String Axiverse}},  {\em Phys. Rev. D} {\bf 81} (2010) 123530, [\href{http://arxiv.org/abs/0905.4720}{{\tt arXiv:0905.4720}}].

\bibitem{Baker:2006ts}
C.~A. Baker et~al., {\it {An Improved experimental limit on the electric dipole moment of the neutron}},  {\em Phys. Rev. Lett.} {\bf 97} (2006) 131801, [\href{http://arxiv.org/abs/hep-ex/0602020}{{\tt hep-ex/0602020}}].

\bibitem{Pendlebury:2015lrz}
J.~M. Pendlebury et~al., {\it {Revised experimental upper limit on the electric dipole moment of the neutron}},  {\em Phys. Rev. D} {\bf 92} (2015), no.~9 092003, [\href{http://arxiv.org/abs/1509.04411}{{\tt arXiv:1509.04411}}].

\bibitem{Hook:2018dlk}
A.~Hook, {\it {TASI Lectures on the Strong CP Problem and Axions}},  {\em PoS} {\bf TASI2018} (2019) 004, [\href{http://arxiv.org/abs/1812.02669}{{\tt arXiv:1812.02669}}].

\bibitem{Peccei:1977hh}
R.~D. Peccei and H.~R. Quinn, {\it {CP Conservation in the Presence of Instantons}},  {\em Phys. Rev. Lett.} {\bf 38} (1977) 1440--1443.

\bibitem{Peccei:1977ur}
R.~D. Peccei and H.~R. Quinn, {\it {Constraints Imposed by CP Conservation in the Presence of Instantons}},  {\em Phys. Rev. D} {\bf 16} (1977) 1791--1797.

\bibitem{Weinberg:1977ma}
S.~Weinberg, {\it {A New Light Boson?}},  {\em Phys. Rev. Lett.} {\bf 40} (1978) 223--226.

\bibitem{Wilczek:1977pj}
F.~Wilczek, {\it {Problem of Strong $P$ and $T$ Invariance in the Presence of Instantons}},  {\em Phys. Rev. Lett.} {\bf 40} (1978) 279--282.

\bibitem{Demirtas:2021gsq}
M.~Demirtas, N.~Gendler, C.~Long, L.~McAllister, and J.~Moritz, {\it {PQ axiverse}},  {\em JHEP} {\bf 06} (2023) 092, [\href{http://arxiv.org/abs/2112.04503}{{\tt arXiv:2112.04503}}].

\bibitem{Kamionkowski:1992mf}
M.~Kamionkowski and J.~March-Russell, {\it {Planck scale physics and the Peccei-Quinn mechanism}},  {\em Phys. Lett. B} {\bf 282} (1992) 137--141, [\href{http://arxiv.org/abs/hep-th/9202003}{{\tt hep-th/9202003}}].

\bibitem{Barr:1992qq}
S.~M. Barr and D.~Seckel, {\it {Planck scale corrections to axion models}},  {\em Phys. Rev. D} {\bf 46} (1992) 539--549.

\bibitem{Ghigna:1992iv}
S.~Ghigna, M.~Lusignoli, and M.~Roncadelli, {\it {Instability of the invisible axion}},  {\em Phys. Lett. B} {\bf 283} (1992) 278--281.

\bibitem{Georgi:1981pu}
H.~M. Georgi, L.~J. Hall, and M.~B. Wise, {\it {Grand Unified Models With an Automatic {Peccei-Quinn} Symmetry}},  {\em Nucl. Phys. B} {\bf 192} (1981) 409--416.

\bibitem{Lazarides:1985bj}
G.~Lazarides, C.~Panagiotakopoulos, and Q.~Shafi, {\it {Phenomenology and Cosmology With Superstrings}},  {\em Phys. Rev. Lett.} {\bf 56} (1986) 432.

\bibitem{Holman:1992us}
R.~Holman, S.~D.~H. Hsu, T.~W. Kephart, E.~W. Kolb, R.~Watkins, and L.~M. Widrow, {\it {Solutions to the strong CP problem in a world with gravity}},  {\em Phys. Lett. B} {\bf 282} (1992) 132--136, [\href{http://arxiv.org/abs/hep-ph/9203206}{{\tt hep-ph/9203206}}].

\bibitem{Ho:2018qur}
S.-Y. Ho, K.~Saikawa, and F.~Takahashi, {\it {Enhanced photon coupling of ALP dark matter adiabatically converted from the QCD axion}},  {\em JCAP} {\bf 10} (2018) 042, [\href{http://arxiv.org/abs/1806.09551}{{\tt arXiv:1806.09551}}].

\bibitem{Chadha-Day:2021uyt}
F.~Chadha-Day, {\it {Axion-like particle oscillations}},  {\em JCAP} {\bf 01} (2022), no.~01 013, [\href{http://arxiv.org/abs/2107.12813}{{\tt arXiv:2107.12813}}].

\bibitem{Foster:2022ajl}
J.~W. Foster, S.~Kumar, B.~R. Safdi, and Y.~Soreq, {\it {Dark Grand Unification in the axiverse: decaying axion dark matter and spontaneous baryogenesis}},  {\em JHEP} {\bf 12} (2022) 119, [\href{http://arxiv.org/abs/2208.10504}{{\tt arXiv:2208.10504}}].

\bibitem{Murai:2023xjn}
K.~Murai, F.~Takahashi, and W.~Yin, {\it {QCD axion: A unique player in the axiverse with mixings}},  {\em Phys. Rev. D} {\bf 108} (2023), no.~3 036020, [\href{http://arxiv.org/abs/2305.18677}{{\tt arXiv:2305.18677}}].

\bibitem{Dvali:1998pa}
G.~R. Dvali and S.~H.~H. Tye, {\it {Brane inflation}},  {\em Phys. Lett. B} {\bf 450} (1999) 72--82, [\href{http://arxiv.org/abs/hep-ph/9812483}{{\tt hep-ph/9812483}}].

\bibitem{Jones:2002cv}
N.~T. Jones, H.~Stoica, and S.~H.~H. Tye, {\it {Brane interaction as the origin of inflation}},  {\em JHEP} {\bf 07} (2002) 051, [\href{http://arxiv.org/abs/hep-th/0203163}{{\tt hep-th/0203163}}].

\bibitem{Frey:2005jk}
A.~R. Frey, A.~Mazumdar, and R.~C. Myers, {\it {Stringy effects during inflation and reheating}},  {\em Phys. Rev. D} {\bf 73} (2006) 026003, [\href{http://arxiv.org/abs/hep-th/0508139}{{\tt hep-th/0508139}}].

\bibitem{Sarangi:2002yt}
S.~Sarangi and S.~H.~H. Tye, {\it {Cosmic string production towards the end of brane inflation}},  {\em Phys. Lett. B} {\bf 536} (2002) 185--192, [\href{http://arxiv.org/abs/hep-th/0204074}{{\tt hep-th/0204074}}].

\bibitem{Bachlechner:2017zpb}
T.~C. Bachlechner, K.~Eckerle, O.~Janssen, and M.~Kleban, {\it {Multiple-axion framework}},  {\em Phys. Rev. D} {\bf 98} (2018), no.~6 061301, [\href{http://arxiv.org/abs/1703.00453}{{\tt arXiv:1703.00453}}].

\bibitem{Hu:2020cga}
D.~Hu, H.-R. Jiang, H.-L. Li, M.-L. Xiao, and J.-H. Yu, {\it {Tale of two- $U(1)$ axion models}},  {\em Phys. Rev. D} {\bf 103} (2021), no.~9 095025, [\href{http://arxiv.org/abs/2009.01452}{{\tt arXiv:2009.01452}}].

\bibitem{Sikivie:1982qv}
P.~Sikivie, {\it {Of Axions, Domain Walls and the Early Universe}},  {\em Phys. Rev. Lett.} {\bf 48} (1982) 1156--1159.

\bibitem{Zeldovich:1974uw}
Y.~B. Zeldovich, I.~Y. Kobzarev, and L.~B. Okun, {\it {Cosmological Consequences of the Spontaneous Breakdown of Discrete Symmetry}},  {\em Zh. Eksp. Teor. Fiz.} {\bf 67} (1974) 3--11.

\bibitem{Hiramatsu:2012sc}
T.~Hiramatsu, M.~Kawasaki, K.~Saikawa, and T.~Sekiguchi, {\it {Axion cosmology with long-lived domain walls}},  {\em JCAP} {\bf 01} (2013) 001, [\href{http://arxiv.org/abs/1207.3166}{{\tt arXiv:1207.3166}}].

\bibitem{Hiramatsu:2010yn}
T.~Hiramatsu, M.~Kawasaki, and K.~Saikawa, {\it {Evolution of String-Wall Networks and Axionic Domain Wall Problem}},  {\em JCAP} {\bf 08} (2011) 030, [\href{http://arxiv.org/abs/1012.4558}{{\tt arXiv:1012.4558}}].

\bibitem{Kibble:1976sj}
T.~W.~B. Kibble, {\it {Topology of Cosmic Domains and Strings}},  {\em J. Phys. A} {\bf 9} (1976) 1387--1398.

\bibitem{Larsson:1996sp}
S.~E. Larsson, S.~Sarkar, and P.~L. White, {\it {Evading the cosmological domain wall problem}},  {\em Phys. Rev. D} {\bf 55} (1997) 5129--5135, [\href{http://arxiv.org/abs/hep-ph/9608319}{{\tt hep-ph/9608319}}].

\bibitem{Saikawa:2017hiv}
K.~Saikawa, {\it {A review of gravitational waves from cosmic domain walls}},  {\em Universe} {\bf 3} (2017), no.~2 40, [\href{http://arxiv.org/abs/1703.02576}{{\tt arXiv:1703.02576}}].

\bibitem{Bai:2023cqj}
Y.~Bai, T.-K. Chen, and M.~Korwar, {\it {QCD-collapsed domain walls: QCD phase transition and gravitational wave spectroscopy}},  {\em JHEP} {\bf 12} (2023) 194, [\href{http://arxiv.org/abs/2306.17160}{{\tt arXiv:2306.17160}}].

\bibitem{Gouttenoire:2025ofv}
Y.~Gouttenoire, S.~F. King, R.~Roshan, X.~Wang, G.~White, and M.~Yamazaki, {\it {Cosmological Consequences of Domain Walls Biased by Quantum Gravity}},  \href{http://arxiv.org/abs/2501.16414}{{\tt arXiv:2501.16414}}.

\bibitem{Vilenkin:1982ks}
A.~Vilenkin and A.~E. Everett, {\it {Cosmic Strings and Domain Walls in Models with Goldstone and PseudoGoldstone Bosons}},  {\em Phys. Rev. Lett.} {\bf 48} (1982) 1867--1870.

\bibitem{Vilenkin:1984ib}
A.~Vilenkin, {\it {Cosmic Strings and Domain Walls}},  {\em Phys. Rept.} {\bf 121} (1985) 263--315.

\bibitem{Davis:1986xc}
R.~L. Davis, {\it {Cosmic Axions from Cosmic Strings}},  {\em Phys. Lett. B} {\bf 180} (1986) 225--230.

\bibitem{Vincent:1996rb}
G.~R. Vincent, M.~Hindmarsh, and M.~Sakellariadou, {\it {Scaling and small scale structure in cosmic string networks}},  {\em Phys. Rev. D} {\bf 56} (1997) 637--646, [\href{http://arxiv.org/abs/astro-ph/9612135}{{\tt astro-ph/9612135}}].

\bibitem{Zhitnitsky:1980tq}
A.~R. Zhitnitsky, {\it {On Possible Suppression of the Axion Hadron Interactions. (In Russian)}},  {\em Sov. J. Nucl. Phys.} {\bf 31} (1980) 260.

\bibitem{Dine:1981rt}
M.~Dine, W.~Fischler, and M.~Srednicki, {\it {A Simple Solution to the Strong CP Problem with a Harmless Axion}},  {\em Phys. Lett. B} {\bf 104} (1981) 199--202.

\bibitem{Iocco:2008va}
F.~Iocco, G.~Mangano, G.~Miele, O.~Pisanti, and P.~D. Serpico, {\it {Primordial Nucleosynthesis: from precision cosmology to fundamental physics}},  {\em Phys. Rept.} {\bf 472} (2009) 1--76, [\href{http://arxiv.org/abs/0809.0631}{{\tt arXiv:0809.0631}}].

\bibitem{Zurek:1985qw}
W.~H. Zurek, {\it {Cosmological Experiments in Superfluid Helium?}},  {\em Nature} {\bf 317} (1985) 505--508.

\bibitem{Copeland:2003bj}
E.~J. Copeland, R.~C. Myers, and J.~Polchinski, {\it {Cosmic F and D strings}},  {\em JHEP} {\bf 06} (2004) 013, [\href{http://arxiv.org/abs/hep-th/0312067}{{\tt hep-th/0312067}}].

\bibitem{Cicoli:2022fzy}
M.~Cicoli, A.~Hebecker, J.~Jaeckel, and M.~Wittner, {\it {Axions in string theory \textemdash{} slaying the Hydra of dark radiation}},  {\em JHEP} {\bf 09} (2022) 198, [\href{http://arxiv.org/abs/2203.08833}{{\tt arXiv:2203.08833}}].

\bibitem{Benabou:2023npn}
J.~N. Benabou, Q.~Bonnefoy, M.~Buschmann, S.~Kumar, and B.~R. Safdi, {\it {The Cosmological Dynamics of String Theory Axion Strings}},  \href{http://arxiv.org/abs/2312.08425}{{\tt arXiv:2312.08425}}.

\bibitem{Bachlechner:2015qja}
T.~C. Bachlechner, C.~Long, and L.~McAllister, {\it {Planckian Axions and the Weak Gravity Conjecture}},  {\em JHEP} {\bf 01} (2016) 091, [\href{http://arxiv.org/abs/1503.07853}{{\tt arXiv:1503.07853}}].

\bibitem{Sikivie:2006ni}
P.~Sikivie, {\it {Axion Cosmology}},  {\em Lect. Notes Phys.} {\bf 741} (2008) 19--50, [\href{http://arxiv.org/abs/astro-ph/0610440}{{\tt astro-ph/0610440}}].

\bibitem{Harlow:2018tng}
D.~Harlow and H.~Ooguri, {\it {Symmetries in quantum field theory and quantum gravity}},  {\em Commun. Math. Phys.} {\bf 383} (2021), no.~3 1669--1804, [\href{http://arxiv.org/abs/1810.05338}{{\tt arXiv:1810.05338}}].

\bibitem{Reece:2023czb}
M.~Reece, {\it {TASI Lectures: (No) Global Symmetries to Axion Physics}},  {\em PoS} {\bf TASI2022} (2024) 008, [\href{http://arxiv.org/abs/2304.08512}{{\tt arXiv:2304.08512}}].

\bibitem{Hiramatsu:2010yz}
T.~Hiramatsu, M.~Kawasaki, and K.~Saikawa, {\it {Gravitational Waves from Collapsing Domain Walls}},  {\em JCAP} {\bf 05} (2010) 032, [\href{http://arxiv.org/abs/1002.1555}{{\tt arXiv:1002.1555}}].

\bibitem{Kawasaki:2011vv}
M.~Kawasaki and K.~Saikawa, {\it {Study of gravitational radiation from cosmic domain walls}},  {\em JCAP} {\bf 09} (2011) 008, [\href{http://arxiv.org/abs/1102.5628}{{\tt arXiv:1102.5628}}].

\bibitem{Hiramatsu:2013qaa}
T.~Hiramatsu, M.~Kawasaki, and K.~Saikawa, {\it {On the estimation of gravitational wave spectrum from cosmic domain walls}},  {\em JCAP} {\bf 02} (2014) 031, [\href{http://arxiv.org/abs/1309.5001}{{\tt arXiv:1309.5001}}].

\bibitem{ZambujalFerreira:2021cte}
R.~Zambujal~Ferreira, A.~Notari, O.~Pujol\`as, and F.~Rompineve, {\it {High Quality QCD Axion at Gravitational Wave Observatories}},  {\em Phys. Rev. Lett.} {\bf 128} (2022), no.~14 141101, [\href{http://arxiv.org/abs/2107.07542}{{\tt arXiv:2107.07542}}].

\bibitem{Kitajima:2023cek}
N.~Kitajima, J.~Lee, K.~Murai, F.~Takahashi, and W.~Yin, {\it {Gravitational waves from domain wall collapse, and application to nanohertz signals with QCD-coupled axions}},  {\em Phys. Lett. B} {\bf 851} (2024) 138586, [\href{http://arxiv.org/abs/2306.17146}{{\tt arXiv:2306.17146}}].

\bibitem{Ferreira:2023jbu}
R.~Z. Ferreira, S.~Gasparotto, T.~Hiramatsu, I.~Obata, and O.~Pujolas, {\it {Axionic defects in the CMB: birefringence and gravitational waves}},  {\em JCAP} {\bf 05} (2024) 066, [\href{http://arxiv.org/abs/2312.14104}{{\tt arXiv:2312.14104}}].

\bibitem{KAGRA:2021kbb}
{\bf KAGRA, Virgo, LIGO Scientific} Collaboration, R.~Abbott et~al., {\it {Upper limits on the isotropic gravitational-wave background from Advanced LIGO and Advanced Virgo\textquoteright{}s third observing run}},  {\em Phys. Rev. D} {\bf 104} (2021), no.~2 022004, [\href{http://arxiv.org/abs/2101.12130}{{\tt arXiv:2101.12130}}].

\bibitem{NANOGrav:2023gor}
{\bf NANOGrav} Collaboration, G.~Agazie et~al., {\it {The NANOGrav 15 yr Data Set: Evidence for a Gravitational-wave Background}},  {\em Astrophys. J. Lett.} {\bf 951} (2023), no.~1 L8, [\href{http://arxiv.org/abs/2306.16213}{{\tt arXiv:2306.16213}}].

\bibitem{Xu:2023wog}
H.~Xu et~al., {\it {Searching for the Nano-Hertz Stochastic Gravitational Wave Background with the Chinese Pulsar Timing Array Data Release I}},  {\em Res. Astron. Astrophys.} {\bf 23} (2023), no.~7 075024, [\href{http://arxiv.org/abs/2306.16216}{{\tt arXiv:2306.16216}}].

\bibitem{Weltman:2018zrl}
A.~Weltman et~al., {\it {Fundamental physics with the Square Kilometre Array}},  {\em Publ. Astron. Soc. Austral.} {\bf 37} (2020) e002, [\href{http://arxiv.org/abs/1810.02680}{{\tt arXiv:1810.02680}}].

\bibitem{LISA:2017pwj}
{\bf LISA} Collaboration, P.~Amaro-Seoane et~al., {\it {Laser Interferometer Space Antenna}},  \href{http://arxiv.org/abs/1702.00786}{{\tt arXiv:1702.00786}}.

\bibitem{Harry:2006fi}
G.~M. Harry, P.~Fritschel, D.~A. Shaddock, W.~Folkner, and E.~S. Phinney, {\it {Laser interferometry for the big bang observer}},  {\em Class. Quant. Grav.} {\bf 23} (2006) 4887--4894. [Erratum: Class.Quant.Grav. 23, 7361 (2006)].

\bibitem{Kawamura:2020pcg}
S.~Kawamura et~al., {\it {Current status of space gravitational wave antenna DECIGO and B-DECIGO}},  {\em PTEP} {\bf 2021} (2021), no.~5 05A105, [\href{http://arxiv.org/abs/2006.13545}{{\tt arXiv:2006.13545}}].

\bibitem{Maggiore:2019uih}
M.~Maggiore et~al., {\it {Science Case for the Einstein Telescope}},  {\em JCAP} {\bf 03} (2020) 050, [\href{http://arxiv.org/abs/1912.02622}{{\tt arXiv:1912.02622}}].

\bibitem{TianQin:2015yph}
{\bf TianQin} Collaboration, J.~Luo et~al., {\it {TianQin: a space-borne gravitational wave detector}},  {\em Class. Quant. Grav.} {\bf 33} (2016), no.~3 035010, [\href{http://arxiv.org/abs/1512.02076}{{\tt arXiv:1512.02076}}].

\bibitem{Ruan:2018tsw}
W.-H. Ruan, Z.-K. Guo, R.-G. Cai, and Y.-Z. Zhang, {\it {Taiji program: Gravitational-wave sources}},  {\em Int. J. Mod. Phys. A} {\bf 35} (2020), no.~17 2050075, [\href{http://arxiv.org/abs/1807.09495}{{\tt arXiv:1807.09495}}].

\bibitem{Reitze:2019iox}
D.~Reitze et~al., {\it {Cosmic Explorer: The U.S. Contribution to Gravitational-Wave Astronomy beyond LIGO}},  {\em Bull. Am. Astron. Soc.} {\bf 51} (2019), no.~7 035, [\href{http://arxiv.org/abs/1907.04833}{{\tt arXiv:1907.04833}}].

\bibitem{Crowder:2005nr}
J.~Crowder and N.~J. Cornish, {\it {Beyond LISA: Exploring future gravitational wave missions}},  {\em Phys. Rev. D} {\bf 72} (2005) 083005, [\href{http://arxiv.org/abs/gr-qc/0506015}{{\tt gr-qc/0506015}}].

\bibitem{Janssen:2014dka}
G.~Janssen et~al., {\it {Gravitational wave astronomy with the SKA}},  {\em PoS} {\bf AASKA14} (2015) 037, [\href{http://arxiv.org/abs/1501.00127}{{\tt arXiv:1501.00127}}].

\bibitem{Widrow:1989fe}
L.~M. Widrow, {\it {The Collapse of Nearly Spherical Domain Walls}},  {\em Phys. Rev. D} {\bf 39} (1989) 3576.

\bibitem{Ferrer:2018uiu}
F.~Ferrer, E.~Masso, G.~Panico, O.~Pujolas, and F.~Rompineve, {\it {Primordial Black Holes from the QCD axion}},  {\em Phys. Rev. Lett.} {\bf 122} (2019), no.~10 101301, [\href{http://arxiv.org/abs/1807.01707}{{\tt arXiv:1807.01707}}].

\bibitem{Gelmini:2022nim}
G.~B. Gelmini, A.~Simpson, and E.~Vitagliano, {\it {Catastrogenesis: DM, GWs, and PBHs from ALP string-wall networks}},  {\em JCAP} {\bf 02} (2023) 031, [\href{http://arxiv.org/abs/2207.07126}{{\tt arXiv:2207.07126}}].

\bibitem{Dunsky:2024zdo}
D.~I. Dunsky and M.~Kongsore, {\it {Primordial Black Holes from Axion Domain Wall Collapse}},  \href{http://arxiv.org/abs/2402.03426}{{\tt arXiv:2402.03426}}.

\bibitem{Kawasaki:2014sqa}
M.~Kawasaki, K.~Saikawa, and T.~Sekiguchi, {\it {Axion dark matter from topological defects}},  {\em Phys. Rev. D} {\bf 91} (2015), no.~6 065014, [\href{http://arxiv.org/abs/1412.0789}{{\tt arXiv:1412.0789}}].

\bibitem{Harigaya:2018ooc}
K.~Harigaya and M.~Kawasaki, {\it {QCD axion dark matter from long-lived domain walls during matter domination}},  {\em Phys. Lett. B} {\bf 782} (2018) 1--5, [\href{http://arxiv.org/abs/1802.00579}{{\tt arXiv:1802.00579}}].

\bibitem{Cline:2024vbd}
J.~M. Cline, C.~Litos, and W.~Xue, {\it {Axion strings from string axions}},  \href{http://arxiv.org/abs/2412.12260}{{\tt arXiv:2412.12260}}.

\bibitem{Sen:2000vx}
A.~Sen, {\it {Non-BPS D-branes in string theory}},  {\em Class. Quant. Grav.} {\bf 17} (2000) 1251--1256.

\bibitem{Dvali:2002fi}
G.~Dvali and A.~Vilenkin, {\it {Solitonic D-branes and brane annihilation}},  {\em Phys. Rev. D} {\bf 67} (2003) 046002, [\href{http://arxiv.org/abs/hep-th/0209217}{{\tt hep-th/0209217}}].

\bibitem{Blanco-Pillado:2005oxi}
J.~J. Blanco-Pillado, G.~Dvali, and M.~Redi, {\it {Cosmic D-strings as axionic D-term strings}},  {\em Phys. Rev. D} {\bf 72} (2005) 105002, [\href{http://arxiv.org/abs/hep-th/0505172}{{\tt hep-th/0505172}}].

\bibitem{Avelino:2005kn}
P.~P. Avelino, C.~J. A.~P. Martins, and J.~C. R.~E. Oliveira, {\it {One-scale model for domain wall network evolution}},  {\em Phys. Rev. D} {\bf 72} (2005) 083506, [\href{http://arxiv.org/abs/hep-ph/0507272}{{\tt hep-ph/0507272}}].

\bibitem{Avelino:2008ve}
P.~P. Avelino, C.~J. A.~P. Martins, J.~Menezes, R.~Menezes, and J.~C. R.~E. Oliveira, {\it {Dynamics of domain wall networks with junctions}},  {\em Phys. Rev. D} {\bf 78} (2008) 103508, [\href{http://arxiv.org/abs/0807.4442}{{\tt arXiv:0807.4442}}].

\bibitem{Kuster:2008zz}
M.~Kuster, G.~Raffelt, and B.~Beltran, eds., {\em {Axions: Theory, cosmology, and experimental searches. Proceedings, 1st Joint ILIAS-CERN-CAST axion training, Geneva, Switzerland, November 30-December 2, 2005}}, vol.~741, 2008.

\bibitem{Hiramatsu:2012gg}
T.~Hiramatsu, M.~Kawasaki, K.~Saikawa, and T.~Sekiguchi, {\it {Production of dark matter axions from collapse of string-wall systems}},  {\em Phys. Rev. D} {\bf 85} (2012) 105020, [\href{http://arxiv.org/abs/1202.5851}{{\tt arXiv:1202.5851}}]. [Erratum: Phys.Rev.D 86, 089902 (2012)].

\bibitem{Yamaguchi:1998gx}
M.~Yamaguchi, M.~Kawasaki, and J.~Yokoyama, {\it {Evolution of axionic strings and spectrum of axions radiated from them}},  {\em Phys. Rev. Lett.} {\bf 82} (1999) 4578--4581, [\href{http://arxiv.org/abs/hep-ph/9811311}{{\tt hep-ph/9811311}}].

\bibitem{Yamaguchi:2002sh}
M.~Yamaguchi and J.~Yokoyama, {\it {Quantitative evolution of global strings from the Lagrangian view point}},  {\em Phys. Rev. D} {\bf 67} (2003) 103514, [\href{http://arxiv.org/abs/hep-ph/0210343}{{\tt hep-ph/0210343}}].

\bibitem{Yamaguchi:2002zv}
M.~Yamaguchi and J.~Yokoyama, {\it {Lagrangian evolution of global strings}},  {\em Phys. Rev. D} {\bf 66} (2002) 121303, [\href{http://arxiv.org/abs/hep-ph/0205308}{{\tt hep-ph/0205308}}].

\bibitem{March-Russell:2021zfq}
J.~March-Russell and H.~Tillim, {\it {Axiverse Strings}},  \href{http://arxiv.org/abs/2109.14637}{{\tt arXiv:2109.14637}}.

\bibitem{Dolan:2017vmn}
M.~J. Dolan, P.~Draper, J.~Kozaczuk, and H.~Patel, {\it {Transplanckian Censorship and Global Cosmic Strings}},  {\em JHEP} {\bf 04} (2017) 133, [\href{http://arxiv.org/abs/1701.05572}{{\tt arXiv:1701.05572}}].

\bibitem{Reece:2018zvv}
M.~Reece, {\it {Photon Masses in the Landscape and the Swampland}},  {\em JHEP} {\bf 07} (2019) 181, [\href{http://arxiv.org/abs/1808.09966}{{\tt arXiv:1808.09966}}].

\bibitem{Lanza:2021udy}
S.~Lanza, F.~Marchesano, L.~Martucci, and I.~Valenzuela, {\it {The EFT stringy viewpoint on large distances}},  {\em JHEP} {\bf 09} (2021) 197, [\href{http://arxiv.org/abs/2104.05726}{{\tt arXiv:2104.05726}}].

\bibitem{Heidenreich:2021yda}
B.~Heidenreich, M.~Reece, and T.~Rudelius, {\it {The Weak Gravity Conjecture and axion strings}},  {\em JHEP} {\bf 11} (2021) 004, [\href{http://arxiv.org/abs/2108.11383}{{\tt arXiv:2108.11383}}].

\bibitem{Blanco-Pillado:2013qja}
J.~J. Blanco-Pillado, K.~D. Olum, and B.~Shlaer, {\it {The number of cosmic string loops}},  {\em Phys. Rev. D} {\bf 89} (2014), no.~2 023512, [\href{http://arxiv.org/abs/1309.6637}{{\tt arXiv:1309.6637}}].

\bibitem{Ringeval:2005kr}
C.~Ringeval, M.~Sakellariadou, and F.~Bouchet, {\it {Cosmological evolution of cosmic string loops}},  {\em JCAP} {\bf 02} (2007) 023, [\href{http://arxiv.org/abs/astro-ph/0511646}{{\tt astro-ph/0511646}}].

\bibitem{Lorenz:2010sm}
L.~Lorenz, C.~Ringeval, and M.~Sakellariadou, {\it {Cosmic string loop distribution on all length scales and at any redshift}},  {\em JCAP} {\bf 10} (2010) 003, [\href{http://arxiv.org/abs/1006.0931}{{\tt arXiv:1006.0931}}].

\bibitem{PhysRevD.31.3052}
T.~Vachaspati and A.~Vilenkin, {\it Gravitational radiation from cosmic strings},  {\em Phys. Rev. D} {\bf 31} (Jun, 1985) 3052--3058.

\bibitem{Allen:1991bk}
B.~Allen and E.~P.~S. Shellard, {\it {Gravitational radiation from cosmic strings}},  {\em Phys. Rev. D} {\bf 45} (1992) 1898--1912.

\bibitem{Allen:1994iq}
B.~Allen and P.~Casper, {\it {A Closed form expression for the gravitational radiation rate from cosmic strings}},  {\em Phys. Rev. D} {\bf 50} (1994) 2496--2518, [\href{http://arxiv.org/abs/gr-qc/9405005}{{\tt gr-qc/9405005}}].

\bibitem{Martins:2016ois}
C.~J. A.~P. Martins, I.~Y. Rybak, A.~Avgoustidis, and E.~P.~S. Shellard, {\it {Extending the velocity-dependent one-scale model for domain walls}},  {\em Phys. Rev. D} {\bf 93} (2016), no.~4 043534, [\href{http://arxiv.org/abs/1602.01322}{{\tt arXiv:1602.01322}}].

\bibitem{Coulson:1995nv}
D.~Coulson, Z.~Lalak, and B.~A. Ovrut, {\it {Biased domain walls}},  {\em Phys. Rev. D} {\bf 53} (1996) 4237--4246.

\bibitem{Ferreira:2024eru}
R.~Z. Ferreira, A.~Notari, O.~Pujol\`as, and F.~Rompineve, {\it {Collapsing domain wall networks: impact on pulsar timing arrays and primordial black holes}},  {\em JCAP} {\bf 06} (2024) 020, [\href{http://arxiv.org/abs/2401.14331}{{\tt arXiv:2401.14331}}].

\bibitem{10.1093/acprof:oso/9780198570745.001.0001}
M.~Maggiore, {\em {Gravitational Waves: Volume 1: Theory and Experiments}}.
\newblock Oxford University Press, 10, 2007.

\bibitem{Caprini:2018mtu}
C.~Caprini and D.~G. Figueroa, {\it {Cosmological Backgrounds of Gravitational Waves}},  {\em Class. Quant. Grav.} {\bf 35} (2018), no.~16 163001, [\href{http://arxiv.org/abs/1801.04268}{{\tt arXiv:1801.04268}}].

\bibitem{Cai:2019cdl}
R.-G. Cai, S.~Pi, and M.~Sasaki, {\it {Universal infrared scaling of gravitational wave background spectra}},  {\em Phys. Rev. D} {\bf 102} (2020), no.~8 083528, [\href{http://arxiv.org/abs/1909.13728}{{\tt arXiv:1909.13728}}].

\bibitem{Saikawa:2018rcs}
K.~Saikawa and S.~Shirai, {\it {Primordial gravitational waves, precisely: The role of thermodynamics in the Standard Model}},  {\em JCAP} {\bf 05} (2018) 035, [\href{http://arxiv.org/abs/1803.01038}{{\tt arXiv:1803.01038}}].

\bibitem{Davis:1985pt}
R.~L. Davis, {\it {Goldstone Bosons in String Models of Galaxy Formation}},  {\em Phys. Rev. D} {\bf 32} (1985) 3172.

\bibitem{Vilenkin:1986ku}
A.~Vilenkin and T.~Vachaspati, {\it {Radiation of Goldstone Bosons From Cosmic Strings}},  {\em Phys. Rev. D} {\bf 35} (1987) 1138.

\bibitem{Auclair:2019wcv}
P.~Auclair et~al., {\it {Probing the gravitational wave background from cosmic strings with LISA}},  {\em JCAP} {\bf 04} (2020) 034, [\href{http://arxiv.org/abs/1909.00819}{{\tt arXiv:1909.00819}}].

\bibitem{Gendler:2023kjt}
N.~Gendler, D.~J.~E. Marsh, L.~McAllister, and J.~Moritz, {\it {Glimmers from the Axiverse}},  \href{http://arxiv.org/abs/2309.13145}{{\tt arXiv:2309.13145}}.

\bibitem{Aloni:2018vki}
D.~Aloni, Y.~Soreq, and M.~Williams, {\it {Coupling QCD-Scale Axionlike Particles to Gluons}},  {\em Phys. Rev. Lett.} {\bf 123} (2019), no.~3 031803, [\href{http://arxiv.org/abs/1811.03474}{{\tt arXiv:1811.03474}}].

\bibitem{Gavela:2023tzu}
B.~Gavela, P.~Qu\'\i{}lez, and M.~Ramos, {\it {The QCD axion sum rule}},  {\em JHEP} {\bf 04} (2024) 056, [\href{http://arxiv.org/abs/2305.15465}{{\tt arXiv:2305.15465}}].

\bibitem{Jaeckel:2017tud}
J.~Jaeckel, P.~C. Malta, and J.~Redondo, {\it {Decay photons from the axionlike particles burst of type II supernovae}},  {\em Phys. Rev. D} {\bf 98} (2018), no.~5 055032, [\href{http://arxiv.org/abs/1702.02964}{{\tt arXiv:1702.02964}}].

\bibitem{Hoof:2022xbe}
S.~Hoof and L.~Schulz, {\it {Updated constraints on axion-like particles from temporal information in supernova SN1987A gamma-ray data}},  {\em JCAP} {\bf 03} (2023) 054, [\href{http://arxiv.org/abs/2212.09764}{{\tt arXiv:2212.09764}}].

\bibitem{Muller:2023vjm}
E.~M\"uller, F.~Calore, P.~Carenza, C.~Eckner, and M.~C.~D. Marsh, {\it {Investigating the gamma-ray burst from decaying MeV-scale axion-like particles produced in supernova explosions}},  {\em JCAP} {\bf 07} (2023) 056, [\href{http://arxiv.org/abs/2304.01060}{{\tt arXiv:2304.01060}}].

\bibitem{Porras-Bedmar:2024uql}
S.~Porras-Bedmar, M.~Meyer, and D.~Horns, {\it {Novel bounds on decaying axionlike particle dark matter from the cosmic background}},  {\em Phys. Rev. D} {\bf 110} (7, 2024) 103501, [\href{http://arxiv.org/abs/2407.10618}{{\tt arXiv:2407.10618}}].

\bibitem{Ayala:2014pea}
A.~Ayala, I.~Dom\'\i{}nguez, M.~Giannotti, A.~Mirizzi, and O.~Straniero, {\it {Revisiting the bound on axion-photon coupling from Globular Clusters}},  {\em Phys. Rev. Lett.} {\bf 113} (2014), no.~19 191302, [\href{http://arxiv.org/abs/1406.6053}{{\tt arXiv:1406.6053}}].

\bibitem{Dolan:2022kul}
M.~J. Dolan, F.~J. Hiskens, and R.~R. Volkas, {\it {Advancing globular cluster constraints on the axion-photon coupling}},  {\em JCAP} {\bf 10} (2022) 096, [\href{http://arxiv.org/abs/2207.03102}{{\tt arXiv:2207.03102}}].

\bibitem{Cadamuro:2011fd}
D.~Cadamuro and J.~Redondo, {\it {Cosmological bounds on pseudo Nambu-Goldstone bosons}},  {\em JCAP} {\bf 02} (2012) 032, [\href{http://arxiv.org/abs/1110.2895}{{\tt arXiv:1110.2895}}].

\bibitem{Turner:1985si}
M.~S. Turner, {\it {Cosmic and Local Mass Density of Invisible Axions}},  {\em Phys. Rev. D} {\bf 33} (1986) 889--896.

\bibitem{Lyth:1991ub}
D.~H. Lyth, {\it {Axions and inflation: Sitting in the vacuum}},  {\em Phys. Rev. D} {\bf 45} (1992) 3394--3404.

\bibitem{Strobl:1994wk}
K.~Strobl and T.~J. Weiler, {\it {Anharmonic evolution of the cosmic axion density spectrum}},  {\em Phys. Rev. D} {\bf 50} (1994) 7690--7702, [\href{http://arxiv.org/abs/astro-ph/9405028}{{\tt astro-ph/9405028}}].

\bibitem{Kobayashi:2013nva}
T.~Kobayashi, R.~Kurematsu, and F.~Takahashi, {\it {Isocurvature Constraints and Anharmonic Effects on QCD Axion Dark Matter}},  {\em JCAP} {\bf 09} (2013) 032, [\href{http://arxiv.org/abs/1304.0922}{{\tt arXiv:1304.0922}}].

\bibitem{Bae:2008ue}
K.~J. Bae, J.-H. Huh, and J.~E. Kim, {\it {Update of axion CDM energy}},  {\em JCAP} {\bf 09} (2008) 005, [\href{http://arxiv.org/abs/0806.0497}{{\tt arXiv:0806.0497}}].

\bibitem{Visinelli:2009zm}
L.~Visinelli and P.~Gondolo, {\it {Dark Matter Axions Revisited}},  {\em Phys. Rev. D} {\bf 80} (2009) 035024, [\href{http://arxiv.org/abs/0903.4377}{{\tt arXiv:0903.4377}}].

\bibitem{Harari:1987ht}
D.~Harari and P.~Sikivie, {\it {On the Evolution of Global Strings in the Early Universe}},  {\em Phys. Lett. B} {\bf 195} (1987) 361--365.

\bibitem{Hagmann:1990mj}
C.~Hagmann and P.~Sikivie, {\it {Computer simulations of the motion and decay of global strings}},  {\em Nucl. Phys. B} {\bf 363} (1991) 247--280.

\bibitem{Davis:1989nj}
R.~L. Davis and E.~P.~S. Shellard, {\it {DO AXIONS NEED INFLATION?}},  {\em Nucl. Phys. B} {\bf 324} (1989) 167--186.

\bibitem{Battye:1993jv}
R.~A. Battye and E.~P.~S. Shellard, {\it {Global string radiation}},  {\em Nucl. Phys. B} {\bf 423} (1994) 260--304, [\href{http://arxiv.org/abs/astro-ph/9311017}{{\tt astro-ph/9311017}}].

\bibitem{Battye:1994au}
R.~A. Battye and E.~P.~S. Shellard, {\it {Axion string constraints}},  {\em Phys. Rev. Lett.} {\bf 73} (1994) 2954--2957, [\href{http://arxiv.org/abs/astro-ph/9403018}{{\tt astro-ph/9403018}}]. [Erratum: Phys.Rev.Lett. 76, 2203--2204 (1996)].

\bibitem{Shellard:1987bv}
E.~P.~S. Shellard, {\it {Cosmic String Interactions}},  {\em Nucl. Phys. B} {\bf 283} (1987) 624--656.

\bibitem{Klaer:2017ond}
V.~B.~. Klaer and G.~D. Moore, {\it {The dark-matter axion mass}},  {\em JCAP} {\bf 11} (2017) 049, [\href{http://arxiv.org/abs/1708.07521}{{\tt arXiv:1708.07521}}].

\bibitem{Vaquero:2018tib}
A.~Vaquero, J.~Redondo, and J.~Stadler, {\it {Early seeds of axion miniclusters}},  {\em JCAP} {\bf 04} (2019) 012, [\href{http://arxiv.org/abs/1809.09241}{{\tt arXiv:1809.09241}}].

\bibitem{Gorghetto:2018myk}
M.~Gorghetto, E.~Hardy, and G.~Villadoro, {\it {Axions from Strings: the Attractive Solution}},  {\em JHEP} {\bf 07} (2018) 151, [\href{http://arxiv.org/abs/1806.04677}{{\tt arXiv:1806.04677}}].

\bibitem{Gorghetto:2020qws}
M.~Gorghetto, E.~Hardy, and G.~Villadoro, {\it {More axions from strings}},  {\em SciPost Phys.} {\bf 10} (2021), no.~2 050, [\href{http://arxiv.org/abs/2007.04990}{{\tt arXiv:2007.04990}}].

\bibitem{Buschmann:2019icd}
M.~Buschmann, J.~W. Foster, and B.~R. Safdi, {\it {Early-Universe Simulations of the Cosmological Axion}},  {\em Phys. Rev. Lett.} {\bf 124} (2020), no.~16 161103, [\href{http://arxiv.org/abs/1906.00967}{{\tt arXiv:1906.00967}}].

\bibitem{Buschmann:2021sdq}
M.~Buschmann, J.~W. Foster, A.~Hook, A.~Peterson, D.~E. Willcox, W.~Zhang, and B.~R. Safdi, {\it {Dark matter from axion strings with adaptive mesh refinement}},  {\em Nature Commun.} {\bf 13} (2022), no.~1 1049, [\href{http://arxiv.org/abs/2108.05368}{{\tt arXiv:2108.05368}}].

\bibitem{Giudice:2000ex}
G.~F. Giudice, E.~W. Kolb, and A.~Riotto, {\it {Largest temperature of the radiation era and its cosmological implications}},  {\em Phys. Rev. D} {\bf 64} (2001) 023508, [\href{http://arxiv.org/abs/hep-ph/0005123}{{\tt hep-ph/0005123}}].

\end{thebibliography}\endgroup

\end{document}